\documentclass[acmlarge,screen,anonymous=false]{acmart}
\AtBeginDocument{%
  \providecommand\BibTeX{{%
    \normalfont B\kern-0.5em{\scshape i\kern-0.25em b}\kern-0.8em\TeX}}}

\setcopyright{acmlicensed}
\acmJournal{IMWUT}
\acmYear{2024} \acmVolume{8} \acmNumber{2} \acmArticle{58} \acmMonth{6}\acmDOI{10.1145/3659620}

\acmPrice{15.00}
\acmISBN{978-1-4503-XXXX-X/18/06}
\usepackage{url}
\usepackage{booktabs}
\usepackage[ruled]{algorithm2e}
\usepackage{algpseudocode}
\usepackage{enumerate}
\usepackage[english]{babel}
\usepackage{blindtext}
\usepackage[caption=false,font=footnotesize,subrefformat=parens]{subfig}
\usepackage{amsmath}

\usepackage{amssymb}
\usepackage{xspace}
\usepackage{multirow}
\usepackage{threeparttable}
\usepackage{float}
\usepackage{flafter}
\usepackage{comment}
\usepackage[skip=12pt]{caption}
\usepackage{enumitem}
\usepackage{placeins}
\def\fig{Fig.\xspace}

\def\tab{Tab.\xspace}

\def\ie{{\textit{i.e.}\xspace}} 
\def\eg{{\textit{e.g.}\xspace}}

\newcommand{\head}[1]{{\noindent \textbf{#1:}}}
\newcommand{\term}[1]{{\textit{#1}}}

\graphicspath{{./figures/}}
\graphicspath{{./pdf/}}
\DeclareGraphicsExtensions{.pdf,.jpeg,.png}

\ifodd 0
\newcommand{\rev}[1]{{\color{black}#1}} %
\newcommand{\revc}[1]{{\color{blue}#1}} %
\newcommand{\com}[1]{\textbf{\color{red}(COMMENT: #1)}} %
\newcommand{\todo}[1]{\textbf{{\color{orange}(TODO: #1)}}}
\else
\newcommand{\rev}[1]{#1}
\newcommand{\revc}[1]{{#1}} %
\newcommand{\com}[1]{}
\newcommand{\todo}[1]{}
\fi

\usepackage{xcolor}
\usepackage[normalem]{ulem}

\def\sysname{\textsc{RFBoost}\xspace}
\def\codeurl{\url{https://github.com/aiot-lab/RFBoost}}
\begin{document}

\title[\sysname]{\sysname: Understanding and Boosting Deep \rev{WiFi} Sensing via Physical Data Augmentation}

\begin{anonsuppress}
	\author{Weiying Hou}
	\email{wyhou@cs.hku.hk}
        \orcid{0009-0006-1264-4649}
	\affiliation{%
		\institution{Department of Computer Science, The University of Hong Kong}
		\city{Hong Kong}
		\country{China}
	}
	
	\author{Chenshu Wu}
	\email{chenshu@cs.hku.hk}
        \orcid{0000-0002-9700-4627}
	\affiliation{%
		\institution{Department of Computer Science, The University of Hong Kong}
		\city{Hong Kong}
		\country{China}
	}
\end{anonsuppress}

\renewcommand{\shortauthors}{Hou and Wu}

\begin{abstract}

Deep learning shows promising performance in wireless sensing. However, deep wireless sensing (DWS) heavily relies on large datasets. Unfortunately, building comprehensive datasets for DWS is difficult and costly, because wireless data depends on environmental factors and cannot be labeled offline. Despite recent advances in few-shot/cross-domain learning, DWS is still facing data scarcity issues. \rev{In this paper, we investigate a distinct perspective of radio data augmentation (RDA) for WiFi sensing and present a data-space solution.} Our key insight is that wireless signals inherently exhibit data diversity, contributing more information to be extracted for DWS. We present \sysname, a simple and effective RDA framework encompassing novel physical data augmentation techniques. We implement \sysname as a plug-and-play module integrated with existing deep models and evaluate it on multiple datasets. Experimental results demonstrate that \sysname achieves remarkable average accuracy improvements of 5.4\% on existing models without additional data collection or model modifications, and the best-boosted performance outperforms 11 state-of-the-art baseline models without RDA. RFBoost pioneers the study of RDA, an important yet currently underexplored building block for DWS, which we expect to become a standard DWS component \rev{of WiFi sensing and beyond}. RFBoost is released at \revc{\codeurl}.

\end{abstract}



\maketitle

\section{Introduction}
\label{sec:intro}

The past decade has witnessed the conceptualization, development and commercialization of wireless sensing, especially WiFi sensing, which has revolutionized how sensing was practiced as well as how WiFi was used \cite{yang2013rssi,ma2019wifi,wu2022wifi}. Many applications have been explored and enabled, including motion detection \cite{wu2015non, zhang2019widetect}, sleep monitoring \cite{zhang2019smars,liu2014wi}, fall detection \cite{palipana2018falldefi,hu2021defall}, gait recognition \cite{wu2020gaitway,wang2016gait}, etc. In recent years, there has been a shift from using signal processing methods to leveraging deep learning methods for wireless sensing \cite{zheng2021enhancing, bhalla2021imu2doppler,gu2021wione,xiao2021onefi,Zhang2021Widar30ZC,cai2020teaching,zhang2018crosssense,jiang2018towards,jiang2020towards}, which we term as \textit{Deep Wireless Sensing} (DWS) for brevity. 
Thanks to the strong ability of feature learning, DWS promises impressive performance compared with conventional methods that usually need complicated handcrafted feature engineering. 

One of the biggest challenges for DWS, however, is building comprehensive wireless datasets for training. 
Large-scale datasets play a critical role in deep learning techniques. The rapid development of Computer Vision (CV) benefits significantly from the wide availability of large-scale public datasets like ImageNet \cite{deng2009imagenet}. 
In DWS, however, it is cumbersome and costly to build a dataset alike, mainly because wireless data (\rev{\eg, WiFi signals}) is difficult to collect and, more importantly, infeasible to label offline. 
Therefore, similar large-scale datasets are so far missing for DWS, which essentially limits the potential and adoption of these techniques. 

It is an exhausting and expensive process to collect and label wireless sensing data. 
On the one hand, as is well recognized, wireless sensing, \rev{especially WiFi sensing}, heavily depends on the environments, placements, subjects, devices, etc. 
To build a comprehensive dataset, one needs to exhaust all these domains and collect abundant data under every setting. 
On the other hand, wireless data cannot be labeled offline, \ie, after the data has been collected. 
While unlabeled images and videos can be helpful by properly labeling them, it is infeasible to achieve the same for wireless data, even with manual efforts, as they are non-perceptual to human eyes. 
To mitigate the data issue, existing works attempt to train DWS models with limited data via techniques like few-shot learning \cite{gu2021wione,xiao2021onefi}, transfer learning \cite{bhalla2021imu2doppler}, generative learning \cite{li2020wi}, cross-modality training \cite{cai2020teaching}, etc. 

\begin{figure}[t]
    \centering
    \includegraphics[width=0.8\textwidth]{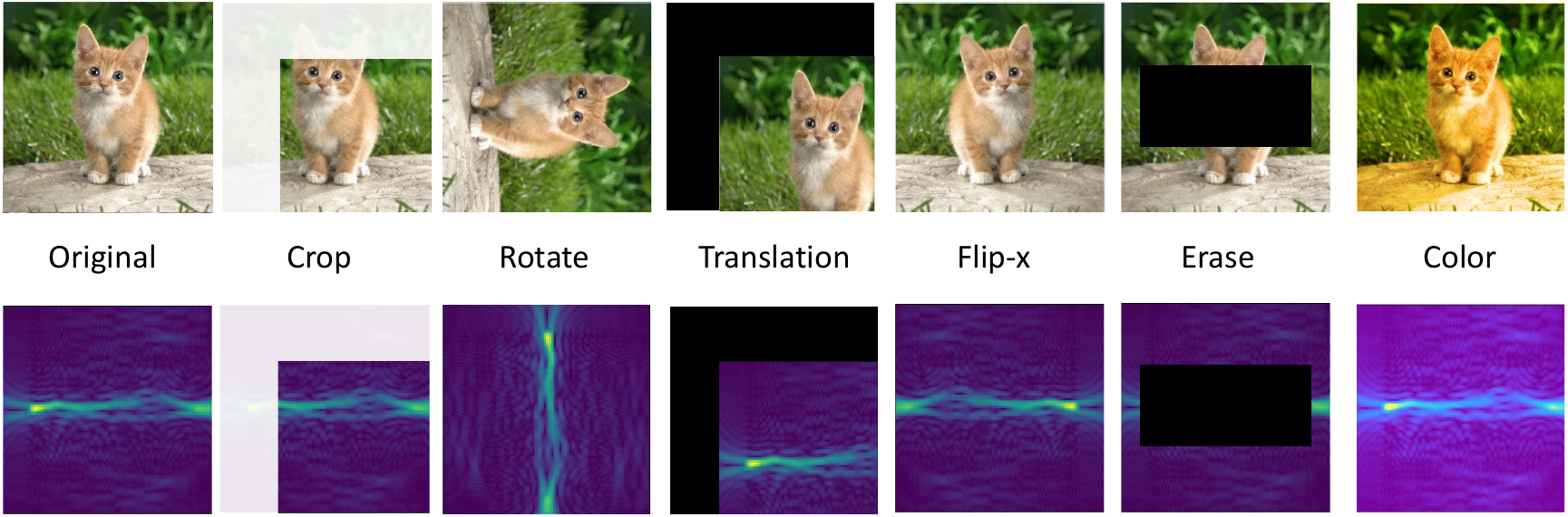}
    \caption{Image data augmentation for images does not apply well to RF spectrograms.}
    \label{fig:vda_rda}
\end{figure}

In this paper, we consider a different perspective from the root of the problem, the training data, and investigate 
\textit{data augmentation} to increase the diversity
of data without collecting more. 
Data augmentation is a set of techniques that artificially inflate the training samples from existing data. 
It has been a common practice and proved effective in CV field for better learning performance. 
The key insight is that more information can be extracted from the original dataset via augmentation. 
Image data augmentation (IDA) for CV is commonly implemented via random transformations, such as flipping, cropping, erasing, rotation, translation, color space transformations, neural style transfer, etc \cite{shorten2019survey}. 
However, as shown in \fig\ref{fig:vda_rda}, these IDA techniques originally designed for images may not be applicable to wireless data due to the fundamental differences in the inherent data structures (See \S\ref{sec:background}). 
New techniques are therefore needed to realize effective data augmentation for wireless sensing data. 
To our best knowledge, no prior work has been dedicated to Radio Data Augmentation (RDA), although the idea has been primarily touched on ad-hoc in the literature \cite{zheng2021more,xiao2021onefi}. 

We target the RDA problem in this work and propose \sysname, the first RDA framework that augments \rev{WiFi sensing data} substantially and effectively. 
The design of \sysname is grounded on the physical properties of WiFi signals. 
Particularly, we investigate physics-inspired data augmentation techniques for DWS by exploiting the inherent data diversity of WiFi signals. 
Signal diversity is well-perceived in wireless communication and sensing and can be observed in all the time, frequency, and space domains. 
Thanks to data diversity, a single sample (\eg, one measurement for a particular gesture/activity) can contribute to multiple independent observations, underpinning a promising opportunity for physical data augmentation (PDA). 
Take frequency diversity as an example. 
Different subcarriers observe diverse responses to human activities and environmental changes, as if an array of cameras capture the same scene with multiple images in one shot. 
While diversity has been well studied, especially for signal processing-based sensing (\eg, subcarrier selection has been a common practice in WiFi sensing \cite{zhang2019widetect,zhang2019smars,wang2015understanding}), it is surprisingly under-exploited in DWS.

To leverage data diversity for effective RDA, we investigate different policies and devise a set of novel techniques to boost the data size. 
We first explore different wireless data representations to understand the performance of DWS, and on this basis, present a DWS framework that takes time-frequency spectrograms as the deep neural networks (DNNs) inputs. 
Then we introduce RDA as a plug-and-play module that can be flexibly added to existing DWS structures. Specifically, it serves as an intermediate processing step between the raw wireless signals and the DNN inputs. 
To leverage frequency diversity, we generate spectrograms on multiple informative subcarriers, and furthermore, we design effective approaches for mixing the spectrograms on different subcarriers. 
Space diversity attributed to multiple antennas, if available, can also be utilized similarly. 
To exploit time diversity, we employ various time windows for spectrogram generation, producing multi-resolution spectrograms for training. 
With all these augmented spectrograms, which are all physically meaningful, \sysname can inflate the training samples considerably and thus improve the learning performance significantly. 
Note that, focusing on data augmentation, \sysname is orthogonal and complementary to research on more advanced DWS models that promise better accuracy or need fewer data. 
It also differs from data synthesis that has been recently explored in the literature \cite{bhalla2021imu2doppler,cai2020teaching,vishwakarma2021simhumalator, Yang2023GANBasedRS}.

We implement \sysname using the PyTorch framework as an open library. 
We validate the RDA effectiveness on three public WiFi sensing datasets for gesture recognition \cite{Zhang2021Widar30ZC}, fall detection \cite{yang2022rethinking}, and human activity recognition \cite{Bocus2022OPERAnetAM}.
We experiment with 11 state-of-the-art models, among which 9 accept spectrogram input (See \tab\ref{tab:models}) and thus can be boosted with \sysname, to examine the performance gains by RDA.
Experimental results demonstrate remarkable performance improvements on all the datasets and models. 
With RDA only, we achieve an average accuracy gain of 5.4\% on 9 existing models without collecting any new data.
Evaluation of the fall detection dataset shows that RDA can \rev{empower small dataset to reach equivalent performance that was previously only achieved by over 4$\times$ of its training data} to achieve the same accuracy of 93\%. 
RDA also proves effective for various DNN models, including standard CV models and those customized for wireless/sensory sensing. 
By introducing RDA to DWS, we believe \sysname will promote practical and deployable DWS and inspire many follow-up research opportunities.

\head{Contributions} 
Our key contributions are summarized as follows: 
\begin{itemize}
    \item We introduce the problem of radio data augmentation to the community as a data-space solution to the data scarcity of DWS. We present an in-depth understanding of DWS by exploring wireless data representations and design \sysname, a simple and effective RDA framework that incorporates a plug-and-play RDA module to boost the training of DWS models. 
    \item We propose a set of physical data augmentation techniques by leveraging the inherent data diversity of \rev{WiFi signals}, which boost training data substantially and effectively in an interpretable way. 
    \item We implement \sysname as an open library and evaluate it on multiple datasets and models. Results show that \sysname augments the training significantly and achieves consistent performance gains across different datasets and models. \sysname's code is released at \codeurl.

\end{itemize}

\section{Preliminaries}
\label{sec:background}

\begin{figure}[t]
    \centering
    \includegraphics[width=0.6\textwidth]{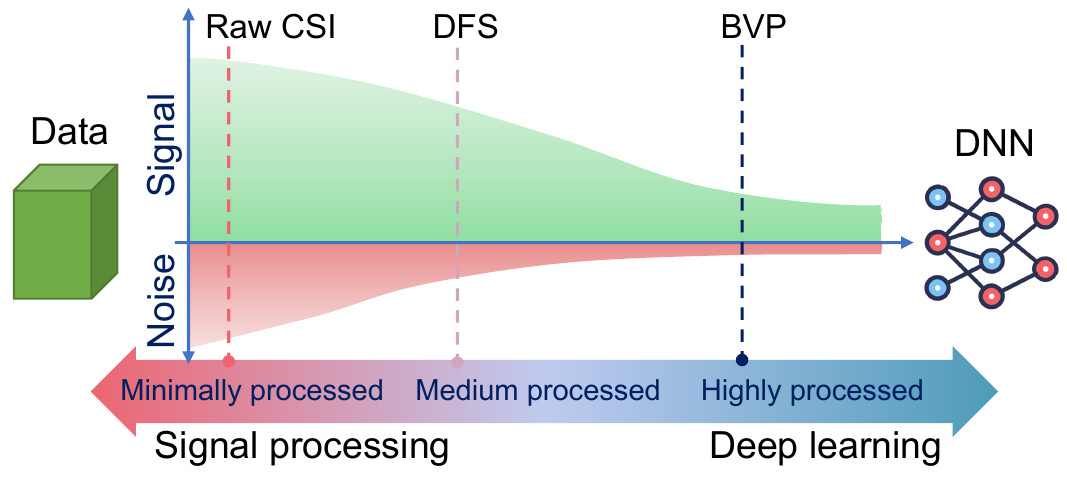}
    \caption{Design choices for data preprocessing in DWS. {\rm Unprocessed or minimally processed data preserves the most information for learning, but suffers from severe noises; Highly processed data denoises the raw signals but also experiences information loss; Medium processed data seems to balance noise reduction and signal reserving for learning.}}
    \label{fig:sp_dl}
\end{figure}

\subsection{Deep Wireless Sensing}
\label{subsec:dws}
Most prior deep wireless sensing works build upon existing neural networks in the CV field, ranging from convolutional neural networks (CNNs) to Generative Adversarial Networks (GANs) and transformers, many of which aim to reduce the need of massive training data \cite{bhalla2021imu2doppler,cai2020teaching,gu2021wione, xiao2021onefi}. 
In the context of WiFi sensing, the source input data is typically the complex Channel State Information (CSI). 
As illustrated in \fig\ref{fig:sp_dl}, without loss of generality, the data representations for the neural network inputs can be roughly categorized as three levels, depending on how much processing is done on the raw data: 

\head{Raw CSI} Some designs argue that DNNs can learn the most from the least preprocessed CSI and thus directly feed in the raw CSI. As CSI is of complex values, the inputs are then usually represented as the CSI amplitude only \cite{zhang2018crosssense}, the amplitude concatenating the phase \cite{ma2018signfi}, and sometimes the I/Q-components \cite{zheng2021more}. 
Raw CSI, or minimally processed radio data, while containing the most information, however, may meanwhile suffer from severe noises. 

\head{Time-frequency spectrograms} Many works, either using signal processing or deep learning, employ time-frequency spectrograms typically generated by Short-Term Fourier Transform (STFT), which embody the Doppler Frequency Shifts (DFSs) for target motion and activities. 
Such spectrograms involve medium efforts of preprocessing and balance between denoising raw data and reserving signals for learning. 
Spectrograms can be represented as images in terms of data format, which can be easily fed into many standard CV models. 
Using time-frequency representations is becoming popular and demonstrated to be effective \cite{yao2019stfnets,li2021units}. 
Considering many subcarriers on WiFi CSI, many works perform Principal Component Analysis (PCA) to reduce the dimension and generate one or a few representative spectrograms \cite{wang2015understanding, qian2018widar2}. 
Alternatively, in DWS, one can also directly use all spectrograms generated on all subcarriers for learning.

\begin{figure}[t]
    \centering
    \includegraphics[width=0.7\textwidth]{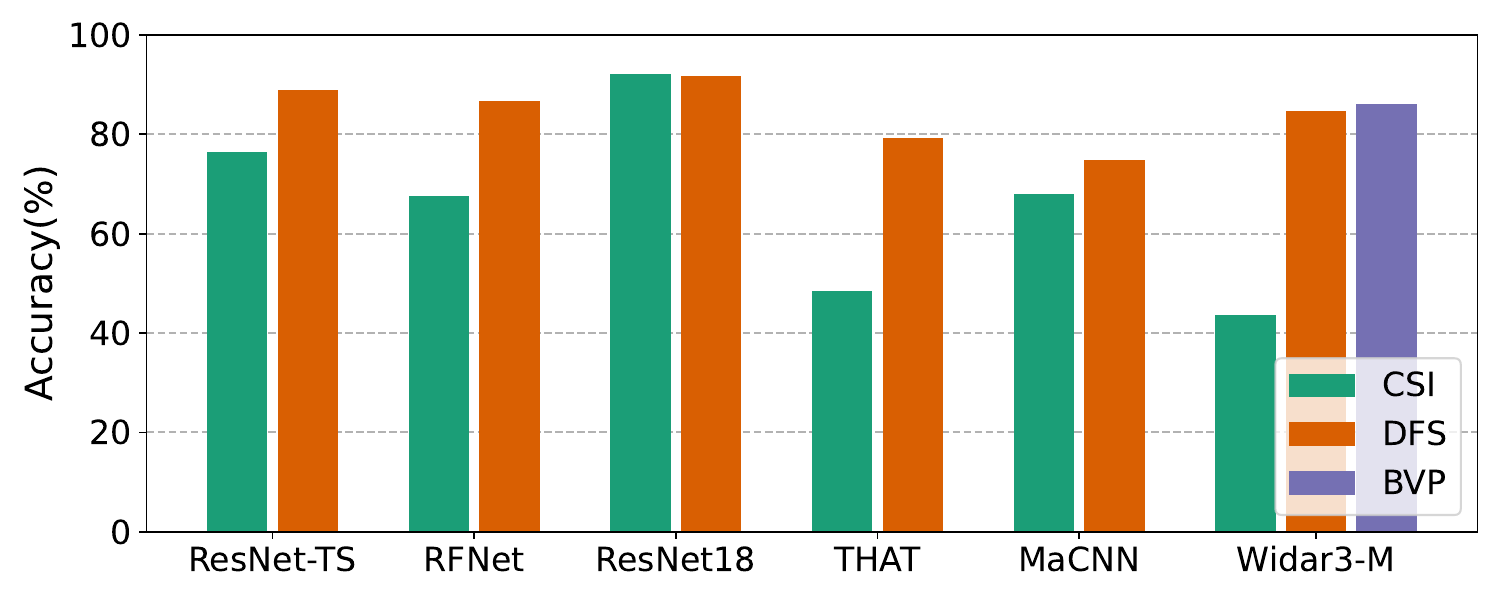}
    \caption {Performance comparison using CSI, DFS, and BVPs as inputs on different models. {\rm DFS inputs outperform raw CSI for most models, and achieve comparable accuracy with BVPs. Note that the original Widar3 model \rev{(Widar3-M)} only accepts BVPs, and is modified here to intake CSI and DFS. \rev{Model description can be found in \tab\ref{tab:models}}}.}
    \label{fig:csi-dfs-bvp}
\end{figure}

\head{Highly-processed information} CSI can be further processed to extract advance information that is supposed to be more explainable and environment-independent, such as \term{Body-coordinate Velocity Profiles} (BVPs) \cite{Zhang2021Widar30ZC} or 2D/3D skeleton \cite{jiang2020towards}, etc. 
While noises can be reduced, a considerable amount of relevant information may be meanwhile lost, confining what DNNs can best learn. 
Furthermore, multiple links are usually compulsory in this case to calculate the desired information, and the computation for the processing can be high \cite{Zhang2021Widar30ZC}. 

In \sysname, we mainly focus on DWS for WiFi sensing with spectrograms as inputs, as the spectrogram representation achieves a good trade-off of the two key components in DWS, \ie, the more explainable but handcrafted preprocessing and the less interpretable yet effective deep learning. 
While conventional wireless sensing frequently optimizes to denoise and reduce the dimension of raw signals, we argue that for DWS, reserving abundant information, even coming with noises, can be beneficial. 

To justify our design choice, we carry out a preliminary experiment based on Widar3 dataset \cite{Zhang2021Widar30ZC}, one of the largest public WiFi sensing datasets thus far, to understand the DWS performance using raw CSI, DFS spectrograms, and BVPs as proposed in Widar3, respectively. 
As shown in \fig\ref{fig:csi-dfs-bvp}, DFS spectrogram inputs significantly outperform raw CSI on most models as expected. 
Surprisingly, DFS inputs also produce comparable performance as the highly-processed and computation-intensive BVPs on the Widar3 model. 
As will be demonstrated later, it will even exceed BVPs when RDA is applied. 
While we agree that the models can be further optimized towards different input formats, the preliminary results with common baselines prove DFS spectrograms as effective data representations for DWS.

\begin{figure*}[t]
    \centering
    \includegraphics[width=\textwidth]{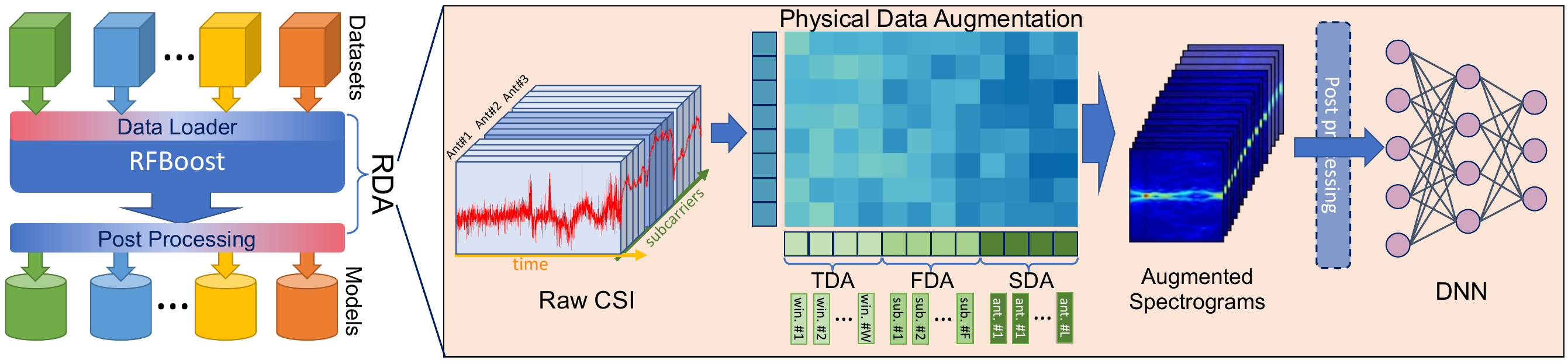}
    \caption{Overview of \sysname design. {\rm RFBoost is designed as a plug-and-play module that can be flexibly integrated with existing models (left). It performs RDA between the raw CSI data and the DNN inputs (right).}}
    \label{fig:overview}
\end{figure*}

\subsection{Data Augmentation}

It is generally recognized that larger datasets promise better deep learning models.  
Data augmentation is thus developed as a data-domain solution to the problem of limited data. %
The target is to enlarge the size and quality of a given training dataset. 
In CV, this is usually done by data warping augmentations which transform existing images with preserved labels by geometric or color transformations, such as cropping, flipping, random erasing, color channel/intensity changes, neural style transfer, etc. 
\fig\ref{fig:vda_rda} illustrates some common operations for image data augmentation, and we refer to \cite{shorten2019survey} for more information. 

These well-studied IDA techniques for image data have performed well on many CV tasks. 
However, they do not perform as well when applied to radio data for augmentation. 
Radio data, even sometimes being represented as images for visualization, has its own underlying structures and physical meanings different from regular images. 
Take time-frequency spectrograms as an example. 
A spectrogram image indicates the frequency response at a certain time, with the horizontal and the vertical dimensions representing the frequency and time, respectively. 
Applying common IDA operations on a spectrogram may change the physical notion and produce unwanted distortions. 
For example, as in \fig\ref{fig:vda_rda}, rotating a spectrogram will reverse the time and frequency dimensions,
flipping it will reverse the time series (\eg, a pull gesture can then become a push gesture), 
and changing the color space does not bring any new information, etc. 
Such operations make little sense to radio data, albeit some of them may still benefit DWS accuracy. 
Although IDA has been extended to other fields like audio and speech processing, data augmentation for DWS has barely been studied so far. 

RDA is an important block of DWS as the problem of data scarcity is particularly critical in DWS due to the difficulties of collecting and labeling wireless data. 
Although there are a few open public datasets available nowadays, data collection for a particular application is usually still required, which renders as one of the biggest hurdles for building a DWS solution. 
Inspired by IDA, \sysname pioneers the study of RDA for various DWS applications and delivers an effective solution. 
Similar to IDA in CV field, RDA is orthogonal to other solutions to handle the issue of limited training data, such as transfer learning \cite{bhalla2021imu2doppler}, pretraining \cite{zheng2021enhancing}, one-shot or zero-shot learning \cite{gu2021wione,xiao2021onefi}, cross-domain learning \cite{Zhang2021Widar30ZC}, cross-modality training \cite{cai2020teaching}, etc., which are out of the scope of this study. 
Also, note that data synthetic approaches like GANs can mitigate the data scarcity problem as well by creating synthetic virtual instances (that are not physical) \cite{zhang2022synthesized}. 
Yet in this work, we keep our focus on \textit{physical data augmentation} (PDA) and do not target synthetic data generation.

\section{\sysname Design}
\label{sec:design}
\subsection{Overview}
\fig\ref{fig:overview} illustrates an overview of the design of \sysname. 
The core of \sysname is the proposed RDA techniques that augment the input data by leveraging time-, frequency-, and space-diversity and output augmented time-frequency spectrograms for further learning. 
We aim to design \sysname as a plug-and-play component that can be easily integrated into many existing and emerging DWS systems. 
Particularly, we envision \sysname to be added into or partly replace the preprocessing pipeline of existing works between raw CSI signals and the DNN inputs. 
Depending on the specific DWS models, certain post-processing can further apply to the augmented outputs of \sysname. 
By default, \sysname only augments data for training, yet we also study the effect of test-time augmentation (\S\ref{sec:tta}).

Our key insight behind \sysname is that the inherent data diversity of wireless signals potentially contributes more information that can be extracted to boost DWS training. 
Such diversity, \eg, frequency diversity, is a generally recognized notion that has been exploited to enhance signal processing performance for wireless sensing. This is usually done by selecting the best or the most representative signal(s) via methods like PCA \cite{yang2013rssi,wang2015understanding,qian2018widar2, palipana2018falldefi} or Maximal Ratio Combining (MRC) \cite{zhang2019smars, wu2020gaitway}. 
While such processing enhances signal-noise-ratio (SNR) and reduces computation for signal processing, it may lead to information loss for deep learning in DWS. 
How to exploit data diversity for RDA is uncharted in DWS. 
Below, we first review the data diversity of WiFi signals and then propose PDA techniques that leverage such signal diversity. 

\subsection{Data Diversity for Physical Data Augmentation}
We first briefly review CSI and explain signal diversity before presenting the proposed RDA techniques. 

CSI reflects the channel of the multipath propagation of wireless signals, which interact with the environments and humans and thus contain environmental information. 
Wireless sensing is then achieved by analyzing the CSI changes due to target movements in the environments. 
In WiFi OFDM systems, CSI is reported from the PHY layer and is estimated on multiple subcarriers, each corresponding to a carrier frequency $f$, as well as each antenna (if there is more than one). 
Therefore, from the data perspective, the time series of CSI measurements can be simply treated as a 3D array $\mathbb{H}_{T\times F\times L}$, where $T$ denotes the total number of WiFi packets (CSI is measured discretely in time on a per packet base), $F$ is the number of subcarriers, and $L=N_{\mathrm{Tx}}\times N_{\mathrm{Rx}}$ denotes the number of links (\ie, the number of Tx antennas) $N_{\mathrm{Tx}}$ multiplying the number of Rx antennas $N_{\mathrm{Rx}}$. 
Each element of $\mathbb{H}_{T\times F\times L}$, denoted as $H(t, f, l)$, represents the CSI measurement at time $t$, subcarrier $f$, and link $l$, which is a complex value with amplitude and phase denoted as follows.
\begin{equation}
    H(t, f, l) = \sum_{m=1}^{M}a_m(t, f, l)\exp(-j 2\pi f\tau_m(t, f, l)) + n(t, f, l),
\end{equation}
where $a_l(t, f, l)$ and $\tau_l(t, f, l)$ are the complex amplitude and propagation delay of the $m$-th multipath component, and $M$ denotes the total number of multipath components. $n(t, f, l)$ stands for the additive thermal noise.  

\begin{figure}[t]
    \centering
    \includegraphics[width=0.8\textwidth]{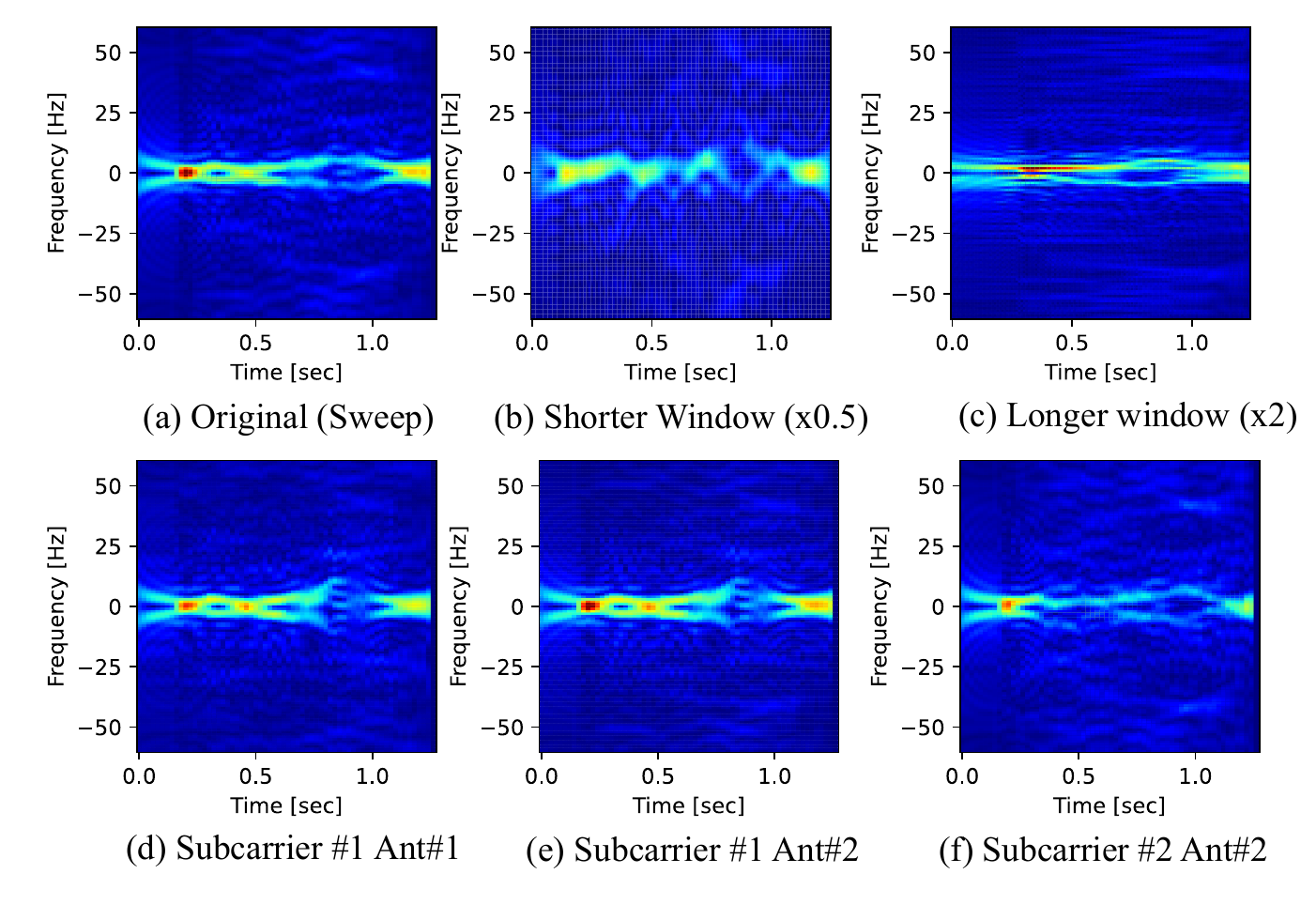}
    \vspace{-0.2in}
    \caption{Data diversity. {\rm For the same activity, the spectrograms using different windows ((b) and (c)), on different antennas ((d) and (e)), on different subcarriers ((e) and (f)), capture different DFS responses and background noise.}}
    \label{fig:diversity}
\end{figure}

\subsubsection{Frequency Diversity} 
CSI reported on different subcarriers exhibit different sensitivities to human activities due to the slightly different underlying carrier frequencies which feature different wavelengths. 
Depending on the specific environments and locations of the target and transceivers, such diversity can be very dominant. 
As in \fig\ref{fig:diversity}, we depict the spectrograms calculated on two subcarriers, a more sensitive subcarrier that captures strong Doppler responses to human motion and a less sensitive subcarrier that contains more noise. 
Although some subcarriers contain larger noise and produce lower SNR spectrograms, they may still contribute information gains that can be learned by DNNs and should not be just discarded.

\subsubsection{Time Diversity}
As mentioned above, to recognize an activity, the raw CSI time series are usually transformed as time-frequency spectrograms via STFT, and the resultant spectrograms basically represent the Doppler Frequency Shifts (DFS) caused by the activity. 
The STFT operation truncates the time series of signals using a sliding time window with a fixed length $w$ and performs FFT for each window, which creates a trade-off between time- and frequency-domain resolutions. 
Using a shorter window reduces the frequency resolution but increases the time responsiveness, and applying a longer window results in the opposite effects. 
Therefore, although high frequency- and time-domain resolutions cannot be achieved simultaneously within the same window, employing multiple windows with diverse lengths will harness the time diversity. 
Spectrograms with shorter windows will better capture fast-changing DFS, while those using longer windows can offer more accurate averaged frequency responses. 
\fig\ref{fig:diversity} portrays spectrograms using different STFT windows. 

\subsubsection{Space Diversity}
\label{sssec:space_di}
Space diversity comes from different antennas, which in principle observe the channel independently at slightly shifted locations. 
Therefore, even on the same subcarrier with the same time window, the generated spectrograms can be considerably different. 
Space diversity is only available when there is more than one antenna forming multiple links (\ie, Tx-Rx antenna pairs). 
Many previous works rely on an antenna array for phased-array processing \cite{qian2018widar2,Zhang2021Widar30ZC}. 
Differently, we exploit multiple antennas as opportunistic diversity for more independent observations.

\vspace{-0.1in}
\subsection{Physical Data Augmentation}
\label{subsec:design_pda}
We now present how to leverage the discussed data diversity for physical data augmentation (PDA) in DWS systems. 

\subsubsection{Frequency-domain Data Augmentation (FDA)}
\label{sssec:design_fda}
A straightforward way to utilize frequency diversity is to generate spectrograms on each subcarrier and add all of them to the training dataset. By doing so, the dataset size will be increased by $F\times L$ times, where $F$ is the total number of subcarriers and $L$ is the number of links. 
This approach may suffer from two issues: 
1) First, depending on the specific WiFi chipsets and driver, there can be tens of or hundreds of subcarriers\footnote{Although a default subcarrier spacing of 312.5 kHz is specified in WiFi standards, different WiFi drivers will report different numbers of subcarriers (\eg, they may downsample or skip some subcarriers in the reported CSI). In addition, the latest IEEE 802.11ax reduces the default subcarrier spacing to 78.125 kHz, adding 4 times more subcarriers than 802.11ac.}. 
Generating spectrograms on each individual subcarriers may not necessarily improve the accuracy, yet will significantly increase the computation. As will be shown in \S\ref{sec:exp}, using all 90 subcarriers on the Widar3 dataset may only produce as good performance as selecting 6 of them. 
2) Second, as subcarriers are continuous over the channel bandwidth, neighboring subcarriers might observe very similar signals. Therefore, spectrograms generated on all subcarriers may suffer from potential redundancy, which should be avoided. 

To optimize FDA, we propose two different policies. For both cases, we consider all the $F$ subcarriers on each link. 

\begin{figure}[t]
    \centering
    \includegraphics[width=0.5\textwidth]{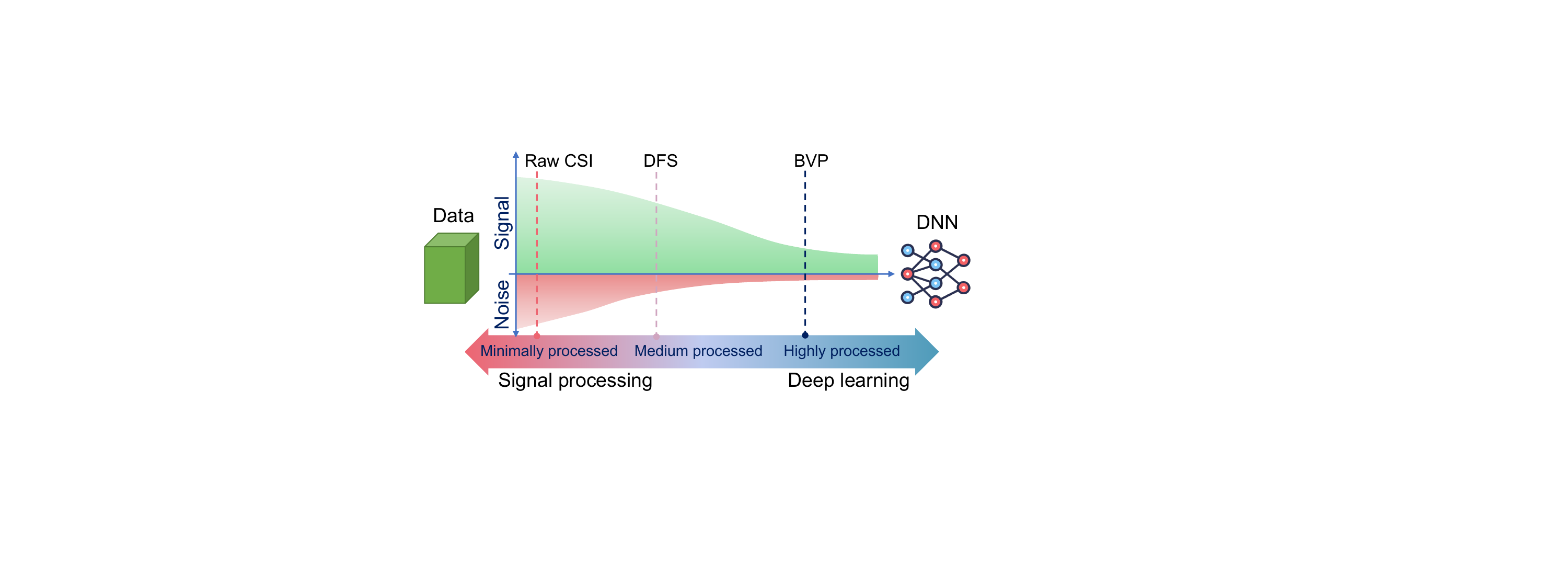}
    \caption{\rev{Individual subcarrier selection when $K=5$. {\rm The frequency band is divided into five sub-bands indicated by the color bars. Subcarriers with the highest motion statistic in each sub-band are selected and marked by red arrows.}}}
    \label{fig:ms-selection}
\end{figure}

(1) \term{Individual Subcarrier Selection (ISS)}. 
Instead of using all the subcarriers, we select only a subset of the most representative subcarriers. 
To do so, we introduce a metric, namely \term{motion statistic} (MS), as the core criteria for selection. 
Motion statistic is a metric for detecting motion and quantifying motion strength using WiFi signals, which is proposed in \cite{zhang2019widetect} based on statistical electromagnetic approaches. 
Specifically, motion statistic is defined as the first sample of the autocorrelation function (ACF) of CSI time series, which can be calculated efficiently. 
Not only being as a motion indicator, but it has also been demonstrated that motion statistic can serve as the equivalent channel gain on each subcarrier, which is useful for sensitive subcarrier selection and optimal subcarrier combining \cite{zhang2019smars, wu2020gaitway}. 

Then our goal is to select a subset of $K$ subcarriers for spectrogram generation such that they experience the highest motion statistics while spanning as broad and uniform as possible over the entire channel bandwidth, referred to as \term{ISS-K}.
As shown in \fig\ref{fig:ms-selection}, we propose a heuristic to select such $K$ subcarriers: Given a fixed $K$, we first divide the entire bandwidth into $K$ uniform sub-bands. 
Then we select the top subcarrier of the highest motion statistic within each sub-band, resulting in $K$ of them. 
Note the value $K$ can be flexibly changed by user input or by certain automatic policy and, typically, is expected to be much smaller than $F\times L$.

(2) \term{Grouped Subcarrier Mixing (GSM)}. As $K\ll F\times L$, many subcarriers will not be used by the above ISS policy only.
To fully exploit frequency diversity, we further propose to group all the subcarriers as $G$ representative clusters and then combine the spectrograms within each cluster as a mixed version for augmentation. 
To make sure the $G$ mixed spectrograms contribute the most information for augmentation, we perform $k$-means clustering to minimize the inter-group similarity and maximize the intra-group similarity, where the similarity is measured as the Euclidean distance between two spectrograms. 
Once the $G$ groups of subcarriers are obtained, we perform Maximal Ratio Combining (MRC) as in \cite{zhang2019smars} on each group to optimally combine all the subcarriers therein. MRC is basically weighted combining, for which we use the motion statistics as the weights here. 
Different from \cite{zhang2019smars} that performs MRC on the ACF signals, we directly use MRC to combine the generated spectrograms on different subcarriers. 
Formally, denote $S(f,t)$ as the spectrogram generated on subcarrier $f$. 
The mixed spectrogram for the subcarriers in group $g$ is then calculated as $S_g(t) = \frac{1}{N_g}\sum_{f\in g}\omega(f,t)S(f,t)$, where $\omega(f,t)$ is the motion statistic for the corresponding subcarrier at time $t$ and $N_g$ denotes the number of subcarriers in the $g$-th group.
\rev{As an alternative, in case of including noisy groups whose average motion statistic indicates relatively low motion among groups
, we can select Top-G groups by sorting the sum of motion statistics in each group instead of using every group as input.}
Similar to the value of $K$, $G$ can also be dynamically changed to fit specific datasets. 
While ISS is not necessarily data augmentation but more about data selection, GSM is true data augmentation since the augmented samples are mixed from multiple spectrograms on individual subcarriers and are thus different from any existing spectrograms generated from a specific subcarrier. 
Additionally, in case multiple antennas are available for space diversity, we can perform FDA separately for each individual antenna for further augmentation.

\subsubsection{Time-domain Data Augmentation (TDA)}
\label{sssec:design_tda}
To leverage time diversity for RDA, we apply $W$ windows, each of a certain length $w_i$, for spectrogram generation. The number of windows and the corresponding window lengths can be given by the user or determined automatically by certain metrics (as will be discussed next). 
In \sysname, by default, we will use $W=3$ windows, one default window, one lengthened window, and one shortened window. \rev{We use a fixed DFT point number ($N_{DFT}$) used by STFT to maintain the uniformity of frequency dimension across different window sizes.}

\begin{figure}[t]
  \begin{minipage}{\textwidth}
      \subfloat[Motion-aware Random Erasing.]{%
          \includegraphics[width=0.49\textwidth]{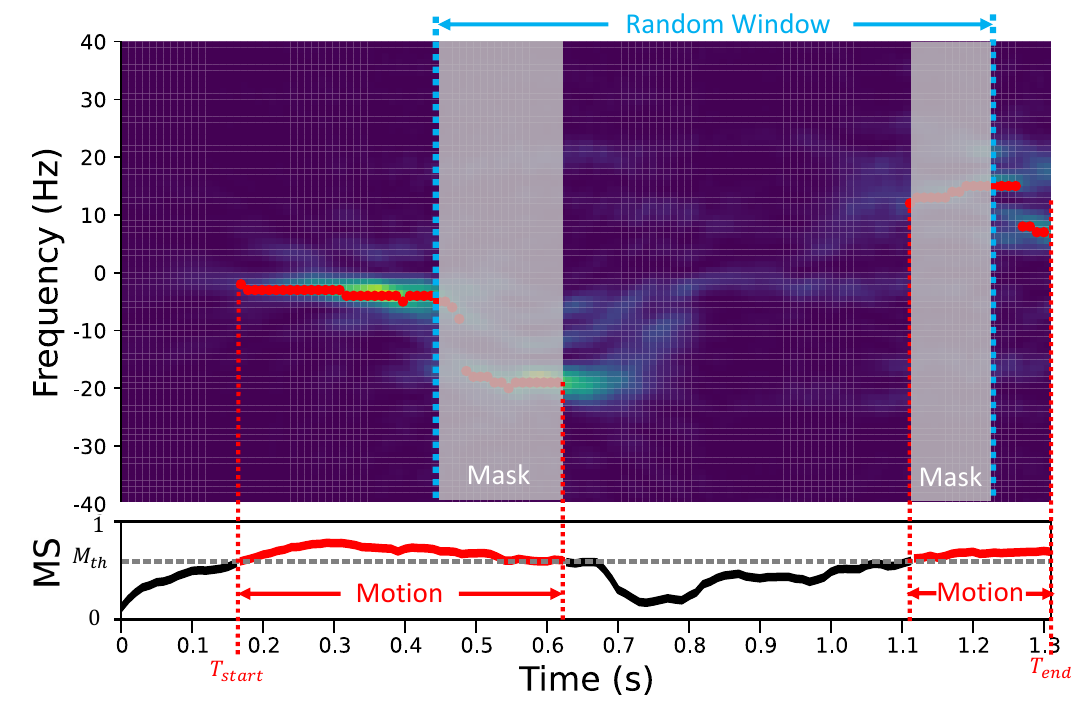}
        \label{subfig:mre}
        }
    \hfill
     \subfloat[\revc{Motion-aware Random Shifting.}]{%
          \includegraphics[width=0.49\textwidth]{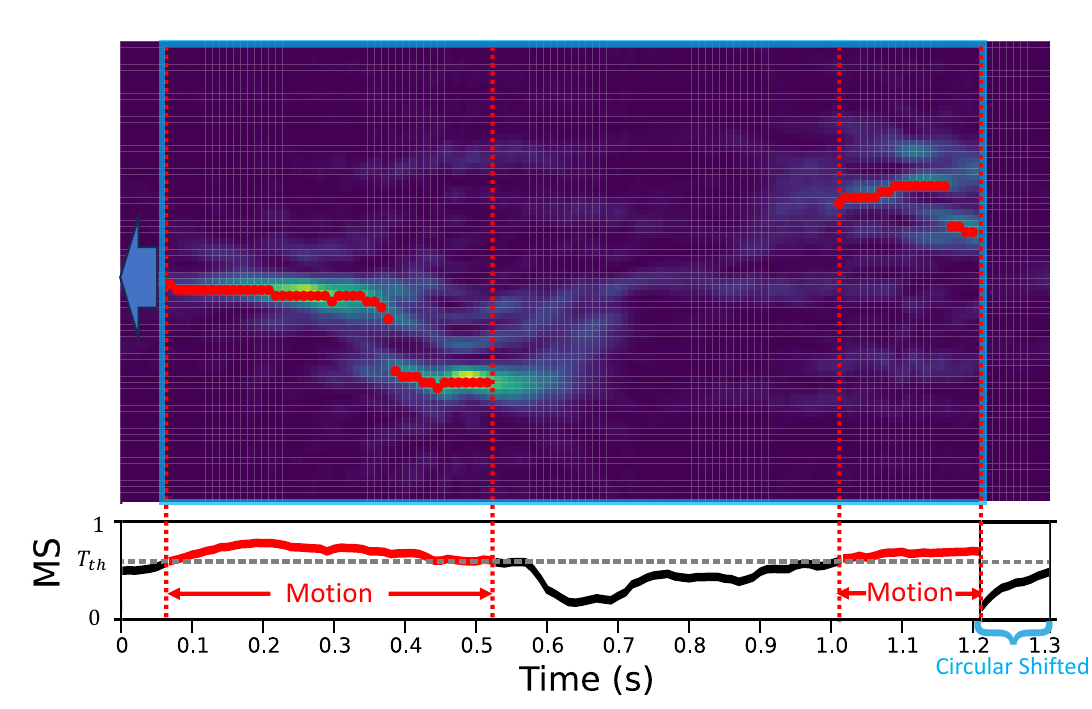}
            \label{subfig:mrs}
    	  }
     \vspace{-0.1in}
     \caption{Motion-aware Data Augmentation. (a) The region of intersection between the \textcolor{red}{motion period} ($MS > T_{th}$) and the \textcolor{blue}{random window} is selectively masked. \rev{Threshold ($M_{th}$) for motion period detection is set to the 2nd quartile point of all measurements of MS in the same environment.} (b) The integrity of the \textcolor{red}{motion period} is maintained in the \textcolor{blue}{blue box} and the marginal "less motion" period is circularly shifted by $-0.1 sec$.}
  
  \end{minipage}
\end{figure}

\subsubsection{Motion-aware Data Augmentation (MDA)}
\label{sssec:mda}
We employ motion statistics to detect the periods of motion as in \cite{zhang2019widetect}. 
With motion awareness, we devise two MDA methods below. 

(1) \term{Motion-aware Random Erasing (MRE)}. 
 Drawing inspiration from object-aware erasing strategies in CV~\cite{zhong2017random, DORNAIKA2023109481}, our \term{Motion-aware Random Erasing} technique strategically places erasure masks at intersections of a randomly chosen window and detected motion-rich region, depicted in \fig\ref{subfig:mre}. This masking process encourages DNNs to emphasize the holistic attributes of the data, thus alleviating overfitting to localized, motion-specific features. It is designed to enhance the model's ability to generalize by obscuring parts of the data that may lead to an excessive focus on minute, potentially non-generalizable details.
 \rev{Different form random time/frequency masking in \cite{Park2019SpecAugmentAS}, our \term{Motion-aware Random Erasing} considers the physical semantics of the spectrogram, which is the motion period.}

(2) \term{Motion-aware Random Shifting (MRS)}. \rev{In contrast to the conventional shifting augmentation used in CV, which often resorts to \textit{padding} to fill the voids created by shifting, our method acknowledges that motion captured in CSI data typically starts and ends with "less motion" period. Utilizing circular shifting in this context does not compromise the physical meaning of the CSI sample while preserving more information than \textit{padding}. Through motion statistics, we detect periods of significant motion, setting the boundaries for circular shifting to maintain the integrity of the "motion" period. For instance, as shown in \fig\ref{subfig:mrs}, since significant motion period is detected within $[T_{start}, T_{end}]$ in this sample with a length of $T$. Then we randomize the shift distance $\tau$, where $\tau\in [-T_{start},  T-T_{end}]$, ensuring the motion period is not circular shifted, and the temporal semantics are maintained while applying MRS.
This approach, similar to how spatial shift in CV reduces overfitting by diversifying a dataset's spatial characteristics, enhances the temporal diversity within the CSI datasets.}

\subsection{Test-time Augmentation}
\label{sec:tta}

\sysname mainly targets training data augmentation to mitigate the data scarcity problem.  
Nevertheless, RDA can also be applied at test time, which has been shown to be effective as well. 
A more robust prediction can be expected by augmenting a test sample in the same way as the training data and then combining the augmented predictions. 
The combination is usually done via a mathematical function or even a learnable neural network \cite{shanmugam2021better}. 
In \sysname, we implement test-time augmentation via a straightforward majority voting method to evaluate the effectiveness.

Test-time augmentation comes at the cost of more computation in the evaluation stage, which translates into an extra delay in model inference. 
Although we do not apply test-time augmentation by default, it can be promising for less time-sensitive tasks in practice.

\section{Implementation}
\label{sec:impl}

\head{Final Dataset Size} 
In the current \sysname, we take a ratio-aware design and keep the final dataset size as an input parameter that can be flexibly determined by users for considerations of specific datasets and different tasks. 
Once it is determined, \ie, given $K$ and $G$ as in \S\ref{sec:design}, \sysname itself can find the best augmentation, \ie, the most informative $K$ subcarriers and the most representative $G$ groups of subcarriers. 
Currently, we only evaluate the performance by empirical choices, but leave it as future work to find the optimal augmentation size and policy automatically. 
We define \term{Augmentation Ratio} (ARatio) as the factor by which an augmentation method increases the size of the original dataset.

\head{Data Length Alignment} 
In most DNN models, the input tensor in the same mini-batch is required to be the same size due to the commonly used fixed size Fully Connected (FC) layer. %
However, in wireless sensing datasets, the length of each sample varies vastly, mainly because it takes various time for different users to repeat different activities. 
\sysname aligns the spectrograms by first padding the CSI time series for STFT and then applying padding, trimming, and/or downsampling to the generated spectrograms, depending on information loss due to such operations.

\head{Online Augmentation} Data augmentation can be done on the fly during training (\ie, online augmentation) or be completed beforehand with augmented data stored (\ie, offline augmentation). 
Online and offline augmentation basically raises a trade-off between training runtime delay and disk I/O performance. 
In our implementation, we design an online approach for augmentation, together with a cache-and-reload mechanism to avoid repeated calculation of temporary results, such as spectrograms for each subcarrier, K-means clustering group, etc.
Nevertheless, the design choice can be flexibly based on the dataset size in practice. 

\head{Software} 
We have developed RFBoost as a PyTorch library, requiring only a customized data loader to handle various raw CSI formats. The enhanced datasets are fully compatible with existing DNN architectures, facilitating easy integration into any PyTorch-based models that process spectrograms or derived formats like \term{Body-coordinate Velocity Profiles}.
Our implementation of the compared baseline methods is partly built upon the released code of UniTS \cite{li2021units}. \sysname code can be found here: \codeurl.

\head{Hardware} 
We train and evaluate \sysname both locally using an NVIDIA A100 and RTX4090 GPU. %
The local A100 GPU is divided into 4 vGPUs using NVIDIA MIG mechanism with a shared cache-base and codebase. 
All the servers share disk Volumes as a codebase and cache-base, respectively.

\section{Experiments}
\label{sec:exp}
We now verify the effectiveness of \sysname via experiments on real-world datasets that are publicly available and on state-of-the-art models. 
We mainly focus on how much improvement RDA can obtain given existing models and datasets.

\begin{table}[t]
\caption{Summary of Compared Baseline Models}
\label{tab:models}
\begin{tabular}{lp{0.4in}c p{0.6\linewidth} p{0.4in}}
\hline
Models                                  & Accepted Inputs && Description &RFBoost-able \\
\toprule
Widar3-2D \cite{Zhang2021Widar30ZC}     &          BVP && Designed for cross-domain gesture recognition using a 2D-CNN+GRU network, with the environment-independent 3D BVPs processed from DFS spectrograms as inputs. & Y           \\
Widar3-1D \cite{Zhang2021Widar30ZC}&          CSI/DFS&& Modified from Widar3-2D by replacing 2D-CNN with 1D-CNN to intake 2D DFS spectrograms. & Y           \\
RF-Net \cite{ding2020rf}                 &         CSI/DFS& & RF-specific meta-learning framework with dual-path base network specially designed for DWS. & Y           \\
THAT \cite{li2021two}                         &          CSI/DFS& & A two convolution augmented transformer model.  & Y           \\
\hline
ResNet \cite{he2016deep}                        &          CSI/DFS& & One of the most used backbone models in CV. & Y           \\
AlexNet \cite{krizhevsky2012imagenet}   &       CSI/DFS& & One classic CV model. & Y           \\
\hline
ResNet-TS \cite{TimeResnet}             &          CSI/DFS && ResNet extended for time series classification. & Y           \\
MaCNN \cite{radu2018multimodal}                      &          CSI/DFS && Multi-modal fusion structure based on CNNs. & Y           \\
LaxCat \cite{hsieh2021explainable}                    &          CSI/DFS && An explainable DNN for time series classification. & Y           \\
SLNet \cite{slnet}      & DFS && A polarized convolutional network designed for spectrograms learning. & Y \\ 
\hline
UniTS \cite{li2021units}                &          CSI && STFT-inspired network for time series sensor data including CSI that achieved the SOTA. & N           \\ 

\bottomrule
\end{tabular}
\end{table}

\subsection{Datasets}
\label{subsec:datasets}
Several WiFi sensing datasets have been made publicly available recently. Our evaluation features three datasets, namely Widar3 dataset \cite{Zhang2021Widar30ZC} for gesture recognition and FallDar \cite{yang2022rethinking} for fall detection, and OPERAnet \cite{Bocus2022OPERAnetAM} for activity recognition.
We choose Widar3 as it is reported the largest WiFi sensing public dataset. 
We are also particularly interested in fall detection, as it is the most difficult to collect real fall data, among many wireless sensing applications. 
It would be exciting to see if \sysname can reduce the need for massive data for fall detection via augmentation. 
More details about each dataset are described below.

\begin{table}[t]
\caption{Descriptions of Widar3-G6 and Widar3-G22 for evaluation.}
\label{tab:widar3_subset}
\footnotesize
\begin{tabular}{lllll}
\hline
                           & Room & Gestures                                                                                & \# of Users & Total \\ \hline
\multirow{3}{*}{Widar3-G6} & 1    & 
\parbox[t]{9cm}{
    1:Push\&Pull;2:Sweep;3:Clap;4:Slide;5:Draw-O(H);6:Draw-Zigzag(H)
} & 9            & 6750          \\
                           & 2    & 
\parbox[t]{9cm}{
    1:Push\&Pull;2:Sweep;3:Clap;4:Slide;5:Draw-O(H);6:Draw-Zigzag(H)
} & 2            & 1500          \\
                           & 3    & 
\parbox[t]{9cm}{
    1:Push\&Pull;2:Sweep;3:Clap;4:Slide;5:Draw-O(H);6:Draw-Zigzag(H)
} & 4            & 3000          \\ \hline
\multirow{2}{*}{Widar3-G22} & 1\&2    & 
\parbox[t]{9cm}{
    0:Slide. 1:Push\&Pull. 2:Clap. 3:Sweep. 4:Draw-Zigzag(Horizontal). 5:Draw-O(Horizontal). 6:Draw-N(Horizontal). 7:Draw-Rectangle(Horizontal). 8:Draw-Triangle(Horizontal). 9:Draw-Zigzag(Vertical). 10:Draw-N(Vertical). 11:Draw-4. 12:Draw-5. 13:Draw-7. 14:Draw-3. 15:Draw-2. 16:Draw-9. 17:Draw-1. 18:Draw-0. 19:Draw-6. 20:Draw-8. 21:Draw-O(Vertical)
} & 2            & 13732          \\ \hline
                           &      &                                                                                         &              &              
\end{tabular}
\end{table}

\subsubsection{Widar3}
\label{sssec:dataset_widar3}
The Widar3 dataset, as released in \cite{Zhang2021Widar30ZC}, encompasses data from 17 participants. These participants are engaged in performing 22 different gestures, 
oriented in 5 different directions towards a transmitter, and stationed at five distinct locations within the range of six receivers, \rev{each equipped with one Intel 5300 network card and placed at different orientations,} across three different environments (rooms). Since not all 22 gestures are performed in every room, we focus on a common subset across all the rooms, which consists of the 6 most frequently performed gestures by 15 users, yielding a total of 11,250 samples, namely Widar3-G6 (or simply Widar3 for short). On Widar3-G6, we will evaluate how \sysname can improve generalizability across unseen devices (\revc{deployment}), environments and users.

\rev{We also use another subset of data from two users who have performed all the 22 gestures (Note that gestures of "Drawing numbers", \ie, Draw-0 to Draw-9, are performed only by two users and in Room\#1 in Widar3 dataset), yielding a total of 13,732 samples, namely Widar3-G22. These subset choices are pivotal in ensuring a comprehensive dataset that captures a diverse range of environmental conditions and participants or gestures to facilitate further cross-domain studies.}
\revc{More details can be found in \tab\ref{tab:widar3_subset}.}

\begin{figure}[t] %
  \centering
  \includegraphics[width=0.35\textwidth]{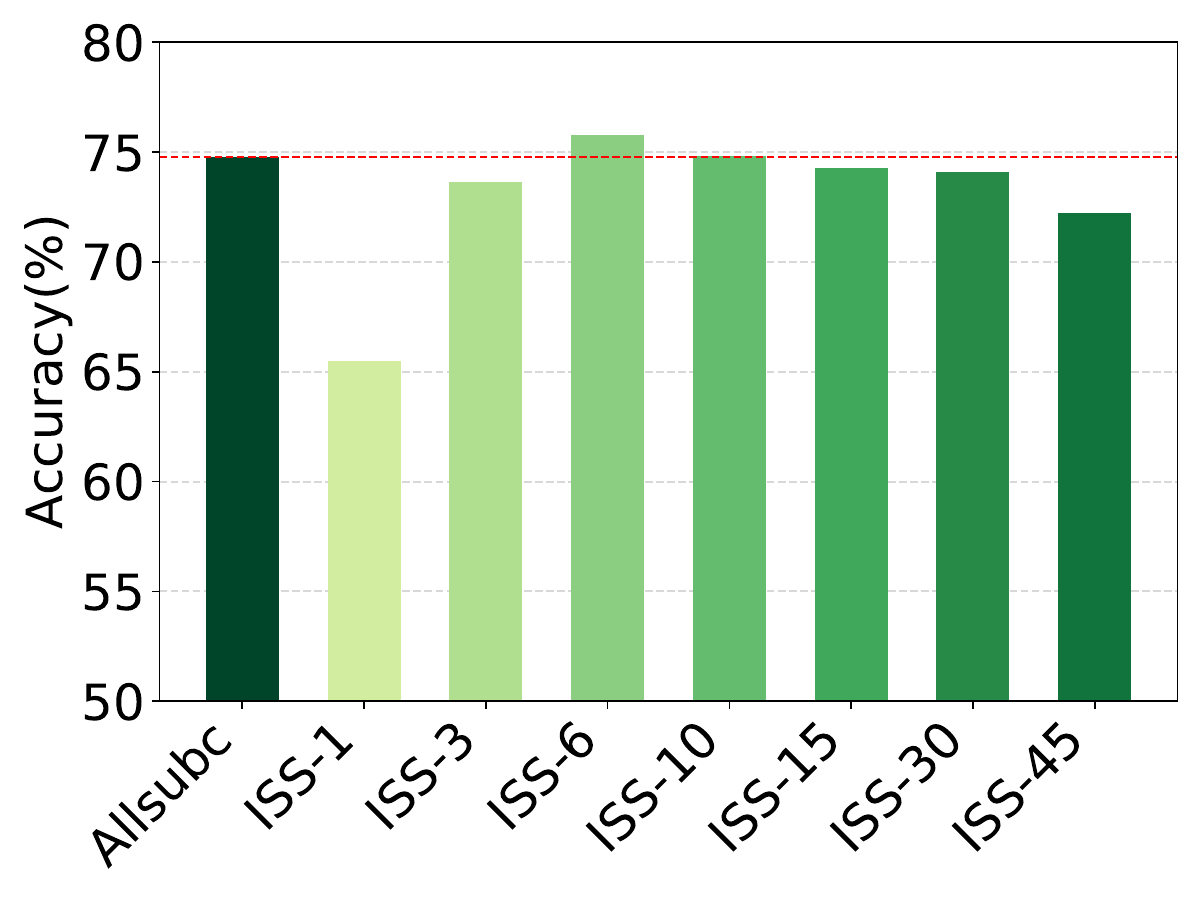}
  \caption{Allsubc v.s. K-subcarriers using proposed Individual Subcarrier Selection (ISS-K, see \S\ref{subsec:design_pda})}
  \label{subfig:ds-ratio}
\end{figure}

\subsubsection{FallDar}
\label{sssec:dataset_fallar}
The FallDar dataset \cite{yang2022rethinking} is collected over six months and includes data from a five-member family, comprising two seniors and three adults, as well as falls simulated with a dummy. The dataset features 442 instances of 6 fall types and approximately 2,000 samples for 21 normal activities. \rev{Among 442 instances of falls, 342 instances are performed by the dummy and the rest 100 instances are done by humans.} CSI data is collected over 30 subcarriers with 3 antennas each. For evaluation, we classify the activities into two categories: fall and non-fall.

\subsubsection{OPERAnet} 
The OPERAnet dataset \cite{Bocus2022OPERAnetAM} is a multi-modal dataset, and we focus on its CSI data collected from two Intel NUC devices for human activity recognition (HAR). This dataset features seven classes of indoor activities recorded over a span of 224 minutes. The activities include body-rotate, lie-down, sit, stand, stand-from-lie, walk, and a no-activity class. Note that OPERAnet dataset demonstrates a marked imbalance in distribution, as depicted in \fig\ref{fig:operanet_gain}. For analytical convenience, we segment these samples into 3.2-second non-overlapping intervals, yielding a total of 5299 samples.

\subsection{Baselines}
Our focus is to evaluate how much relative performance improvements can be gained by \sysname's RDA, given existing datasets and models, rather than optimizing the absolute accuracy by, \eg, pursuing sophisticated model design, which is out of our scope. 
Bearing this in mind, we consider the following categories of 11 baselines:
\begin{itemize}[leftmargin=*]
    \item Models adapted/customized for DWS, such as RF-Net \cite{ding2020rf}, Widar3 \cite{Zhang2021Widar30ZC} model and its variants, THAT \cite{li2021two}, SLNet \cite{slnet}, etc. We will examine how much RDA can further improve models designed for wireless sensing. 
    \item Common CV models, such as ResNet \cite{he2016deep}, AlexNet \cite{deng2009imagenet}, etc. We want to see the performance, with RDA, on standard CV models that can be employed with little knowledge of wireless sensing and no model changes towards DWS.
    \item Advanced models optimized for spatial-temporal learning with sensory time series, including ResNet-TS \cite{TimeResnet}, MaCNN \cite{radu2018multimodal}, LaxCat \cite{hsieh2021explainable}, UniTS \cite{li2021units}, SLNet \cite{slnet}, etc. These models are mostly designed for time series of general sensor data, including WiFi CSI data. 
\end{itemize}
\tab\ref{tab:models} summarizes the baseline models we compared. 
We integrate \sysname into all these models for evaluation except for UniTS which only takes raw CSI as inputs.
Some of these original models do not intake DFS spectrograms. Nevertheless, we can still feed DFS spectrograms as long as they accept 2D tensors as inputs. 
Again, we are particularly interested in understanding how RDA can enhance the performance of models that are not optimized for spectrograms or even not for wireless data.

\begin{figure}[t]
  \begin{minipage}{\textwidth}
      \subfloat[]{%
          \includegraphics[width=0.42\textwidth]{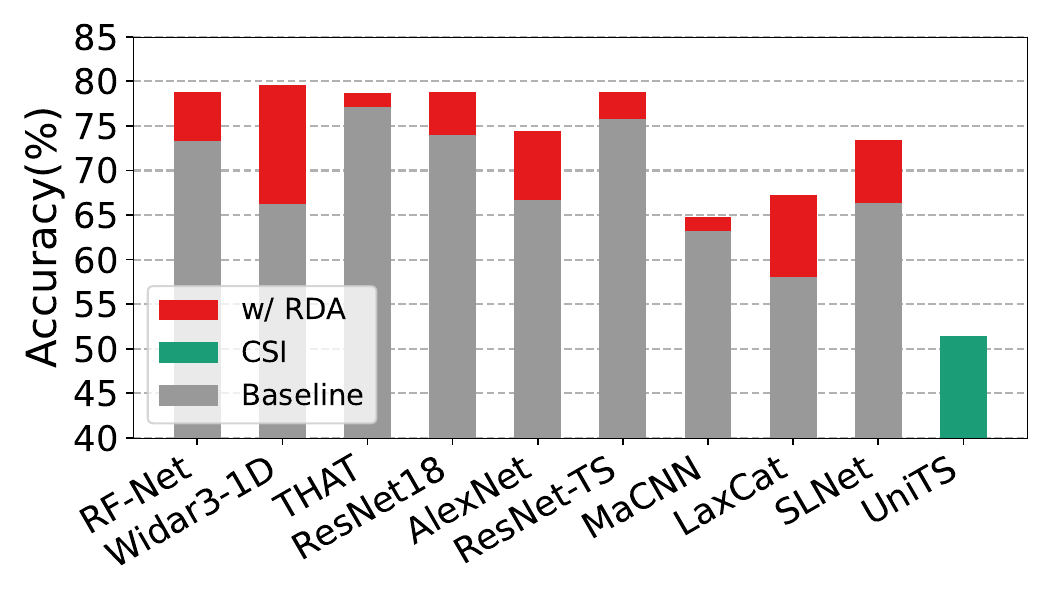}
        \label{fig:widar3-sota-x}
        }
    \hfill
     \subfloat[]{%
          \includegraphics[width=0.42\textwidth]{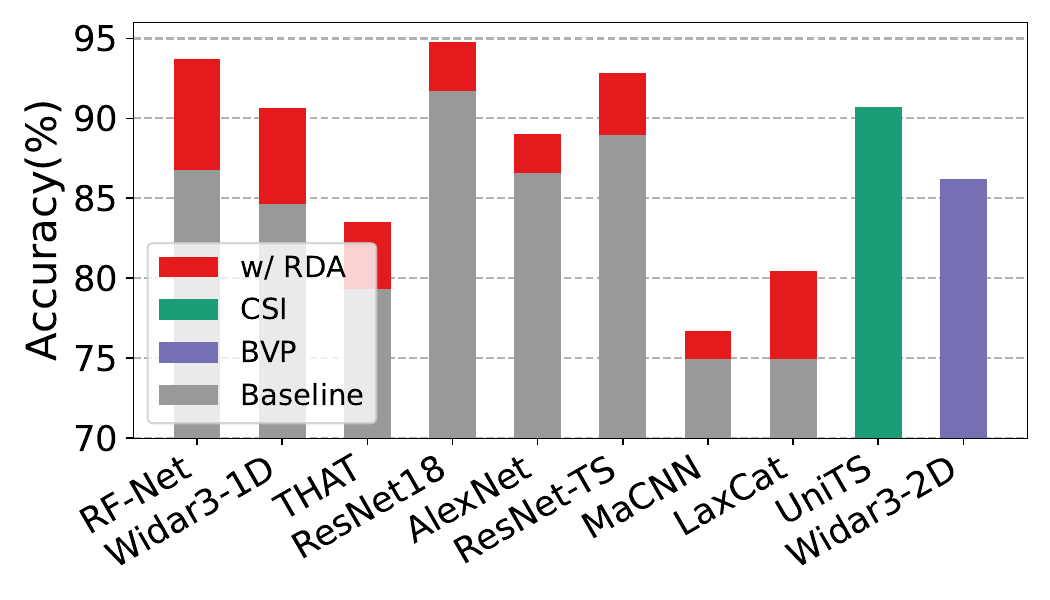}
            \label{fig:widar3-sota-in}
    	  }
     \hfill
     \subfloat[]{%
        \label{subfig:widar3-input}
          \includegraphics[width=0.12\textwidth]{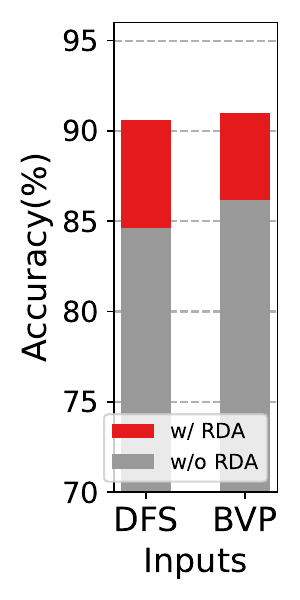}
    	  }
    	\label{fig:widar3-input}
     \vspace{-0.1in}
         \caption{RDA performance on Widar3 dataset in cross-Rx (a) and in-domain (b,c) evaluations and RDA on BVP compared with DFS using Widar3-2D (c).}
  
  \end{minipage}
\end{figure}

To ensure fair comparisons between model performance w/ and w/o RDA on the dataset, uniform training configurations are employed for training on each model. This includes a consistent batch size of 256, learning rate of 0.001, and a standardized optimizer selection; AdamW is the default, with the exception of SLNet which utilizes RMSprop used in its original work. In alignment with our focus on assessing the data augmentation efficacy of \sysname, we guarantee convergence of each model under these training conditions by setting a maximum number of epochs and an early stopping criterion based on validation loss plateauing for a predetermined number of patience epochs.

\begin{figure}[t]
  \begin{minipage}{0.33\textwidth}
    \centering
    \includegraphics[width=1\textwidth]{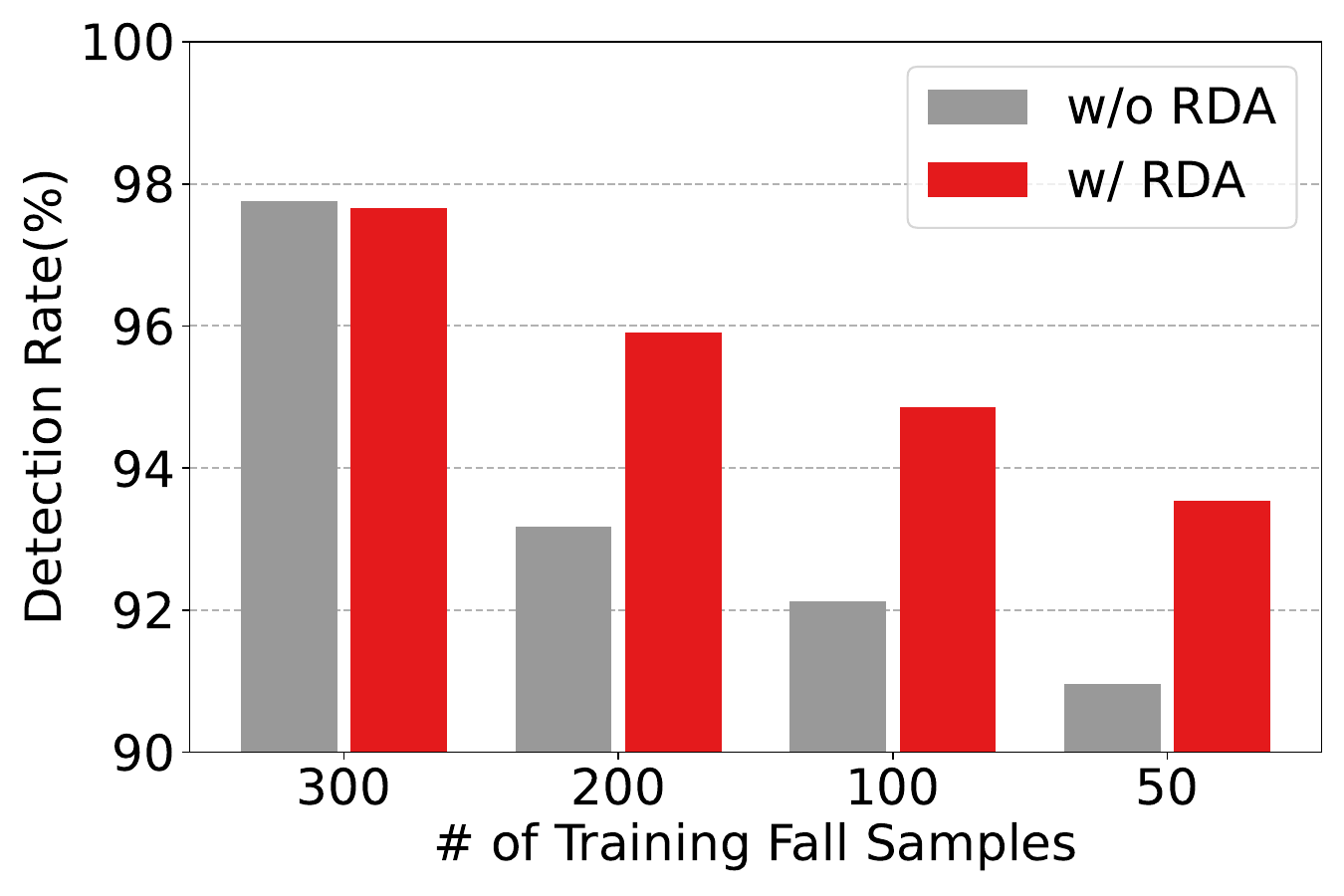}
    \caption{Detection Rate on FallDar for fall detection.}

        \label{fig:falldar_result-dr}
  \end{minipage}
  \hfill
  \begin{minipage}{0.32\textwidth}
    \centering
    \includegraphics[width=1\textwidth]{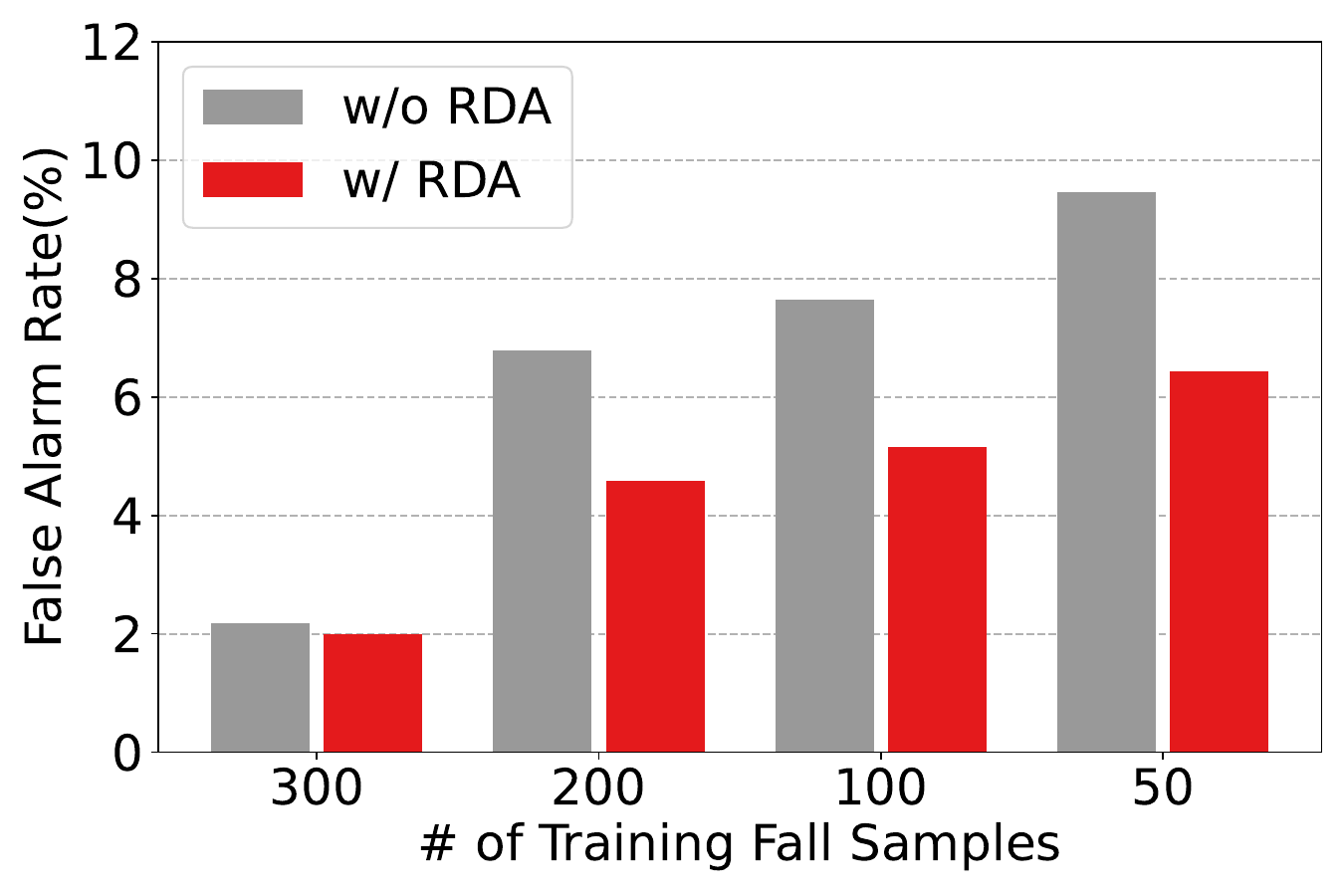}
    \caption{False Alarm Rate on FallDar for fall detection.}
    \label{fig:falldar_result-far}
  \end{minipage}
    \begin{minipage}{0.33\textwidth}
    \centering
    \includegraphics[width=\textwidth]{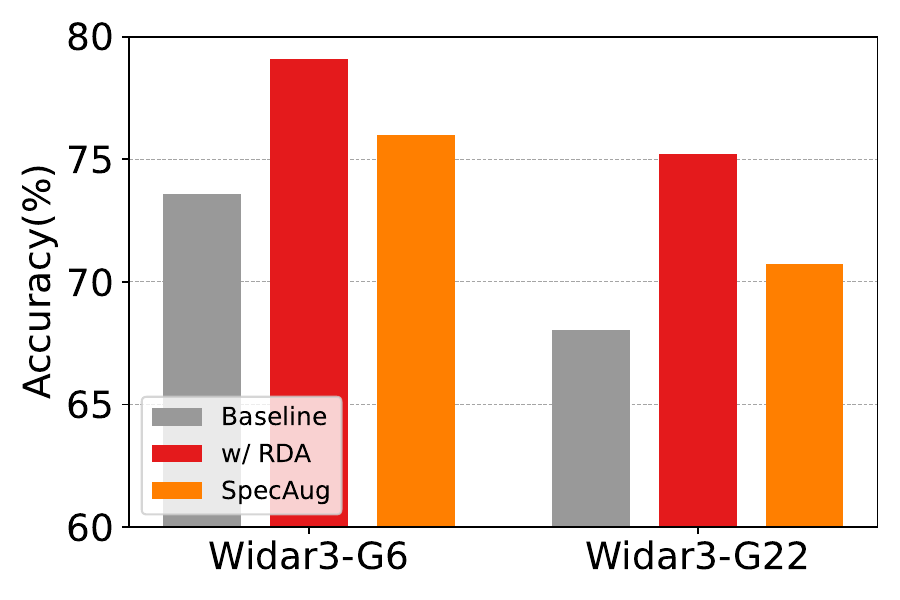}
    \vspace{-0.2in}
    \caption{\rev{Cross-Rx evaluation on Widar3-G6 and Widar3-G22 datasets.}}
    \label{fig:widar3-gesture-6vs22}
  \end{minipage}
\end{figure}

\subsection{\sysname Effectiveness}
\label{subsec:effectiveness}

We mainly focus our evaluation on the relative performance gains attributed to \sysname, especially in terms of generalizability for cross-domain scenarios.

\subsubsection{Evaluation Settings}
In principle, all subcarriers should undergo augmentation. However, doing so is computationally expensive. For instance, training a single augmented epoch on the Widar3 dataset takes over 96 minutes on our NVIDIA A100 GPU server. To accelerate evaluation, we compare using all subcarriers against a selected subset. As shown in \fig\ref{subfig:ds-ratio}, a subset of just six selected subcarriers nearly matches the performance of using all. Implementing RDA on this subset saves 90 minutes per epoch and maintains comparable performance. Consequently, for computational efficiency, our evaluations use only six subcarriers selected by ISS-6 for augmentation, while keeping the baseline as using all subcarriers, \ie, 90 subcarriers for Widar3 dataset.
We first highlight the overall effectiveness by using an augmentation ratio (ARatio) of 9 (2$\times$ TDA and 6$\times$ MDA plus 1 $\times$ FDA\rev{; ARatio calculated in reference to the size of the subset}). For fair comparisons, all comparative studies are conducted in the same environment and evaluation settings.

\subsubsection{Gesture Recognition} 
\label{sssec:eval_ges}
We mainly focus on \term{cross-Rx} evaluation on Widar3-G6 dataset in our effectiveness evaluation. Specifically, Widar3 dataset reports CSI from six Rx devices positioned in various orientations. As a result, cross-Rx evaluation assesses the model's generalization capabilities across both different devices and their orientations/locations. For training, we select CSI from 3 of the 6 Rx while CSI from the other 3 Rx is discarded. In the test phase, we only test the model on the unseen Rx devices in the training phase.
We use cross-Rx as the default setting unless otherwise specified.

We train and evaluate all the applicable baseline models on the Widar3 dataset with and without RDA, respectively. 
As shown in \fig\ref{fig:widar3-sota-x}, while the absolute accuracy is dependent on the specific model, \sysname consistently improves the performance of all models by 5.4\% on average, with the highest gain of 13.4\% as well as the highest absolute accuracy of 79.63\% being achieved on Widar3-1D. 
We also compare the original Widar3 model (denoted as Widar3-2D) \cite{Zhang2021Widar30ZC}, which uses a 2D-CNN+GRU structure and takes \term{Body-coordinate Velocity Profiles} (BVPs) as inputs. 

We adapted the original Widar3-2D implementation to the PyTorch framework, noting that it processes 3D BVPs across the x-y spatial plane and time axis. To accommodate 2D DFS spectrograms and raw CSI, which encompass the frequency and time axes, we developed Widar3-1D.
Comparing \fig\ref{fig:widar3-sota-in} and \fig\ref{subfig:widar3-input}, Widar3-2D, even using BVPs, does not yield the best performance compared with other models, especially those augmented by \sysname. 
Although Widar3-2D using BVPs does perform better than the adapted Widar3-1D for DFS spectrograms (and much better than using raw CSI as in \fig\ref{fig:csi-dfs-bvp}), the performance difference becomes negligible when both are augmented by \sysname (As BVPs are calculated from DFS, they can also be augmented by \sysname). 
The results validate our insights in \S\ref{sec:background} that the highly processed information such as BVPs may lose considerable information and thus prevent the best possible learning performance. Additionally, with the raw CSI as input, UniTS performs among the best in in-domain evaluation while dropping to 51\% in cross-Rx scenarios, which validates our insights that raw CSI has bad generalization ability due to severe noises.
Further considering the intensive overhead to compute BVPs, we argue that BVPs may not be the best choice for deep wireless sensing. 
Instead, DFS or other forms of spectrograms appear to be a superior representation, especially with RDA.

\begin{figure}[t]
  \begin{minipage}{\textwidth}
      \subfloat[Accuracy gains w/ RDA]{%
          \includegraphics[width=0.5\textwidth]{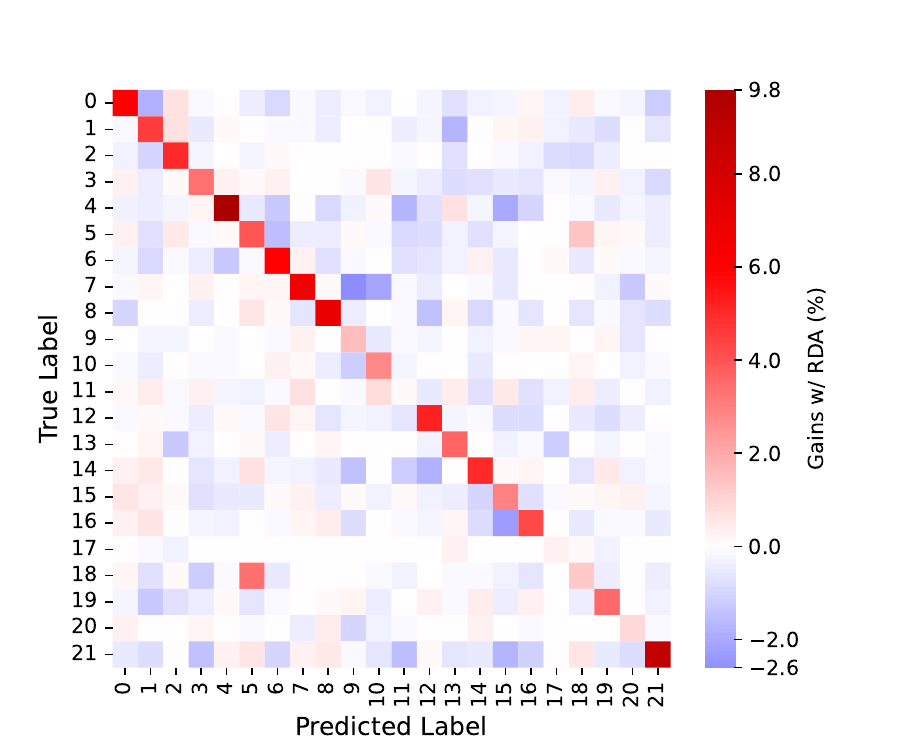}
        \label{subfig:widar3_cm_gesture22_rda}
        }
    \hfill
     \subfloat[Accuracy gains w/ SpecAugment]{%
          \includegraphics[width=0.5\textwidth]{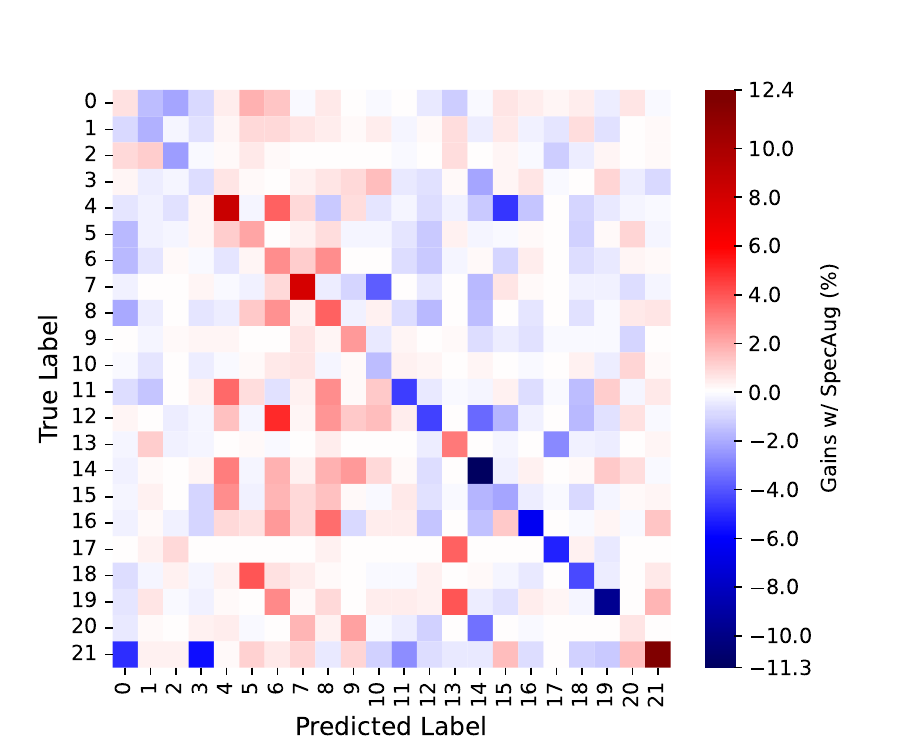}
            \label{subfig:widar3_cm_gesture22_spec}
    	  }
     \hfill
         \caption{\rev{Confusion matrix for performance gains (compared with confusion matrix of baseline) on Widar3-G22 dataset w/ RDA (a) or w/ SpecAugment (b). Both confusion matrices share the same color map with min\&max values marked at color bars.}}
         \label{fig:widar3_cm_gesture22}
  
  \end{minipage}
\end{figure}

\rev{We also conduct experiments on Widar3-G22, which involves all 22 gestures, to further evaluate the performance of distinguishing similar gesture patterns. As the result shown in \fig\ref{fig:widar3-gesture-6vs22}, RDA achieves a 7.16\% accuracy gain compared with baseline using all subcarriers, compared with the gain of 5.52\% on Widar-G6 dataset. As the number of gestures increases, there demands more training samples for models to learn the temporal features, thus, RDA introduces much more performance gain. As depicted in \fig\ref{subfig:widar3_cm_gesture22_rda}, w/ RDA, most gestures are classified more accurately, \eg, Gesture7 (Draw-Rectangle(Horizontal)), which is often mistaken as Gesture9 (Draw-Zigzag(Vertical)) or Gesture10 (Draw-N(Vertical)), receives an improvement of accuracy of 7\% and reaches up to 85\%. We also note that one pair of gestures become even harder to distinguish, which is Gesture18 (Draw-O) and Gesture5 (Draw-0) as they are extremely similar gestures according to the gesture illustration demos released on Widar3 website\footnote{https://ieee-dataport.org/open-access/widar-30-wifi-based-activity-recognition-dataset}. Although the error between Gesture5 and Gesture18 is increased, the overall performance of these two gestures are still improved by 4\% and 1.2\% w/ RDA since RDA improves model’s ability to differentiate between these gestures with other patterns.
}

\subsubsection{Fall Detection}
For fall detection on the FallDar dataset, rather than repeating the evaluation on Widar3, we instead attempt to understand how many training samples for falls are (minimally) needed to achieve a reasonable performance. 
To answer this, we conduct a different experiment by gradually excluding the training samples for falls and comparing the performance by applying RDA. 
Specifically, there are around 442 fall samples in total. We use RF-Net as the baseline and train it with gradually reduced amounts of fall samples from 300 to 50 for training. %
\fig\ref{fig:falldar_result-dr} and \fig\ref{fig:falldar_result-far} depict the results. 
As expected, the performance, in terms of detection rate and false alarm rate, degrades with respect to shrunk training samples of falls, regardless of applying RDA or not. 
However, \sysname improves the performance in any case, with about 2.7\% gain in detection rate and on average 2.6\% decrease in false alarm rate. 
More importantly, by applying RDA, we achieve the same performance using only 50 training samples of falls as that of using 200 samples without RDA, remarkably reducing the required amount by 4$\times$.
\begin{figure*}[t]
  \begin{minipage}{\textwidth}
      \subfloat[Without RDA]{%
          \includegraphics[width=0.48\textwidth]{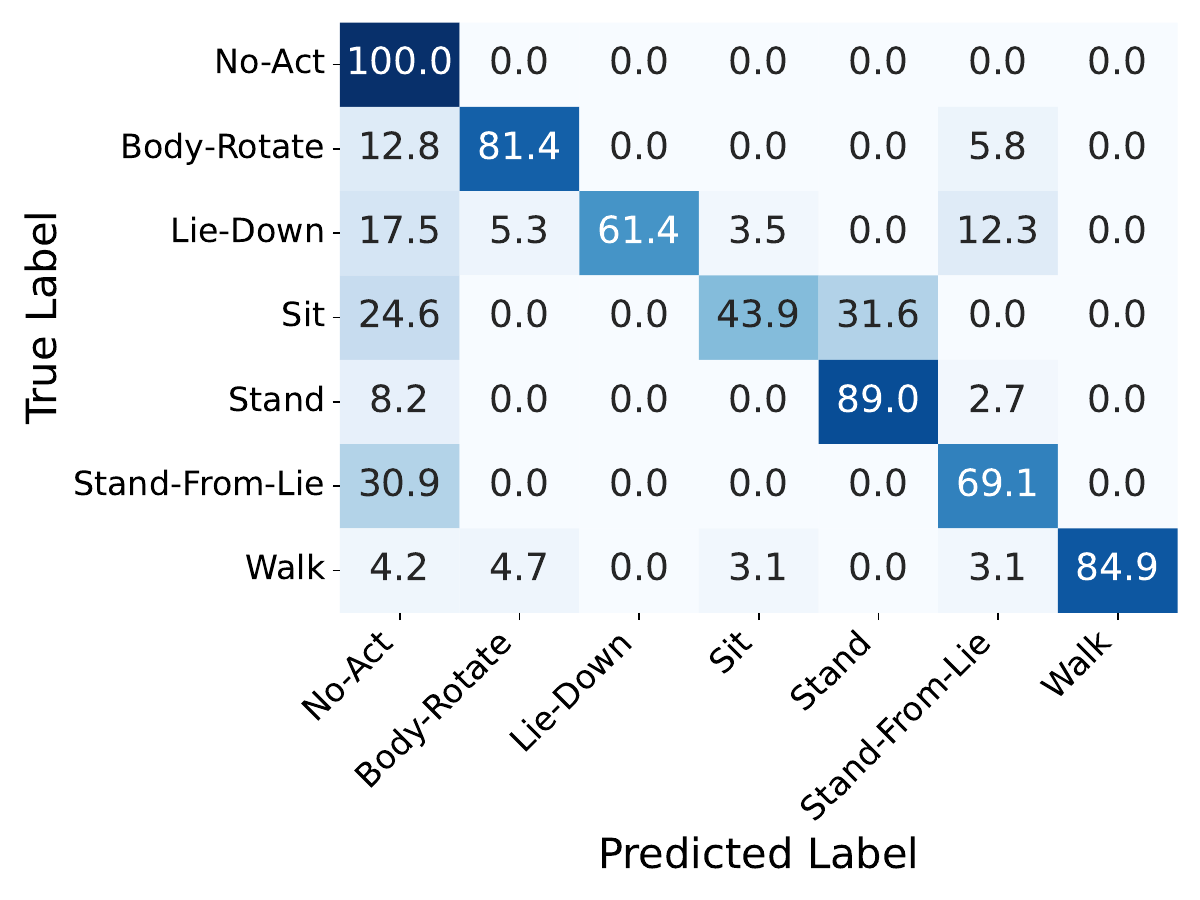}
        \label{subfig:operanet_baseline}
        }
    \hfill
     \subfloat[With RDA]{%
          \includegraphics[width=0.48\textwidth]{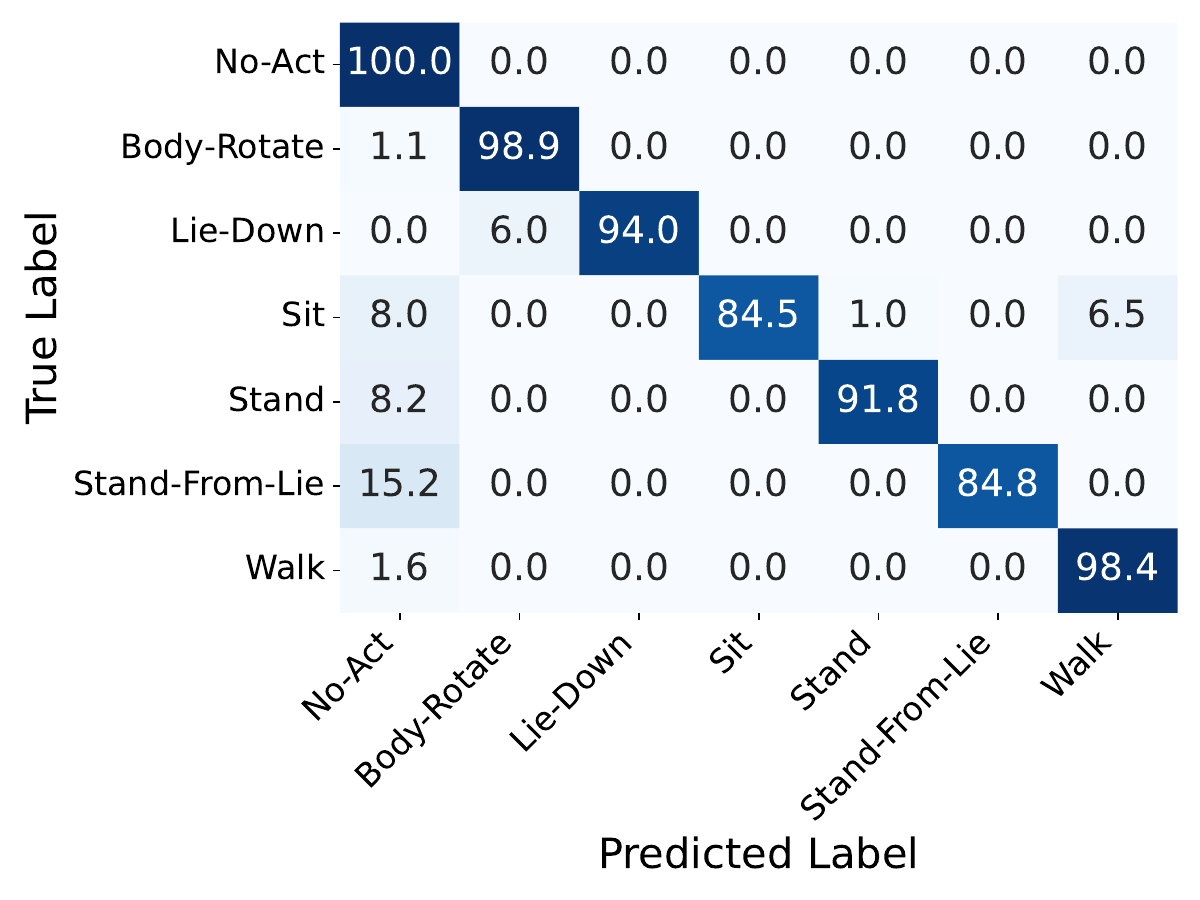}
            \label{subfig:operanet_rda}
    	  }
     \hfill
     \vspace{-0.1in}
     \caption{The confusion matrices on OPERAnet dataset for human activity recognition.}
     \label{fig:operanet_performance}
     \vspace{0.3in}
  \end{minipage}
\end{figure*}

\begin{figure}[t]
    \centering
    \includegraphics[width=0.5\textwidth]{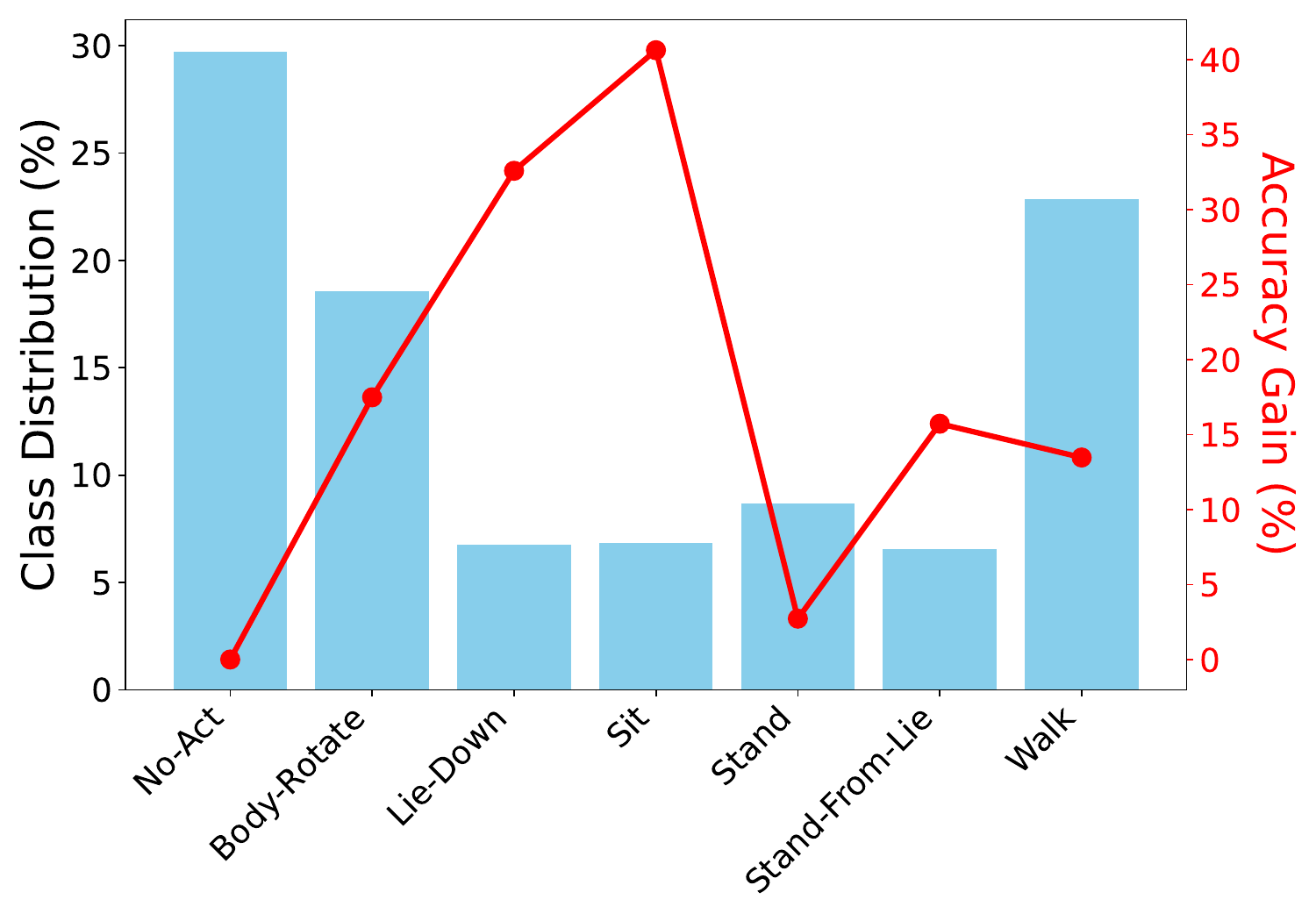}
    \vspace{-0.2in}
    \caption{Class distribution on OPERAnet dataset and accuracy gain after applying RDA.}
    \label{fig:operanet_gain}
\end{figure}

\subsubsection{Human Activity Recognition}
In assessing \sysname's efficacy in addressing dataset imbalances, we turn our attention to the OPERAnet dataset, which is characterized by significant imbalanced label distribution (See \fig\ref{fig:operanet_gain}).
Such imbalances mirror the real-world scenarios in human activity recognition (HAR), where the duration and frequency of activities naturally vary. Typically, longer-duration activities like walking are over-represented, whereas critical but short-duration activities like falling are underrepresented. However, it is often these infrequent activities that are of greatest interest and importance.
To investigate whether \sysname can effectively mitigate the challenges posed by an imbalanced dataset, we employ RF-Net as the baseline model, which achieves an accuracy of 83.69\% using the original dataset. Implementing \sysname results in an enhanced overall accuracy of 96.26\% the overall accuracy. As depicted in \fig\ref{fig:operanet_performance}, the skewed distribution of the dataset adversely impacts the performance of less represented classes, such as 'sit', which constitutes a mere 6\% of the total dataset. 
Upon the application of RDA, there is a notable improvement observed across all classes, averaging a 17\% increase in performance. Significantly, the underrepresented classes exhibit a more substantial elevation in accuracy, effectively addressing the issue of limited data availability. For instance, the 'sit' class witnesses an exceptional increment in performance, surging by 40.6\%.

\begin{figure}[t]
  \begin{minipage}{0.49\textwidth}
    \centering
    \includegraphics[width=1\textwidth]{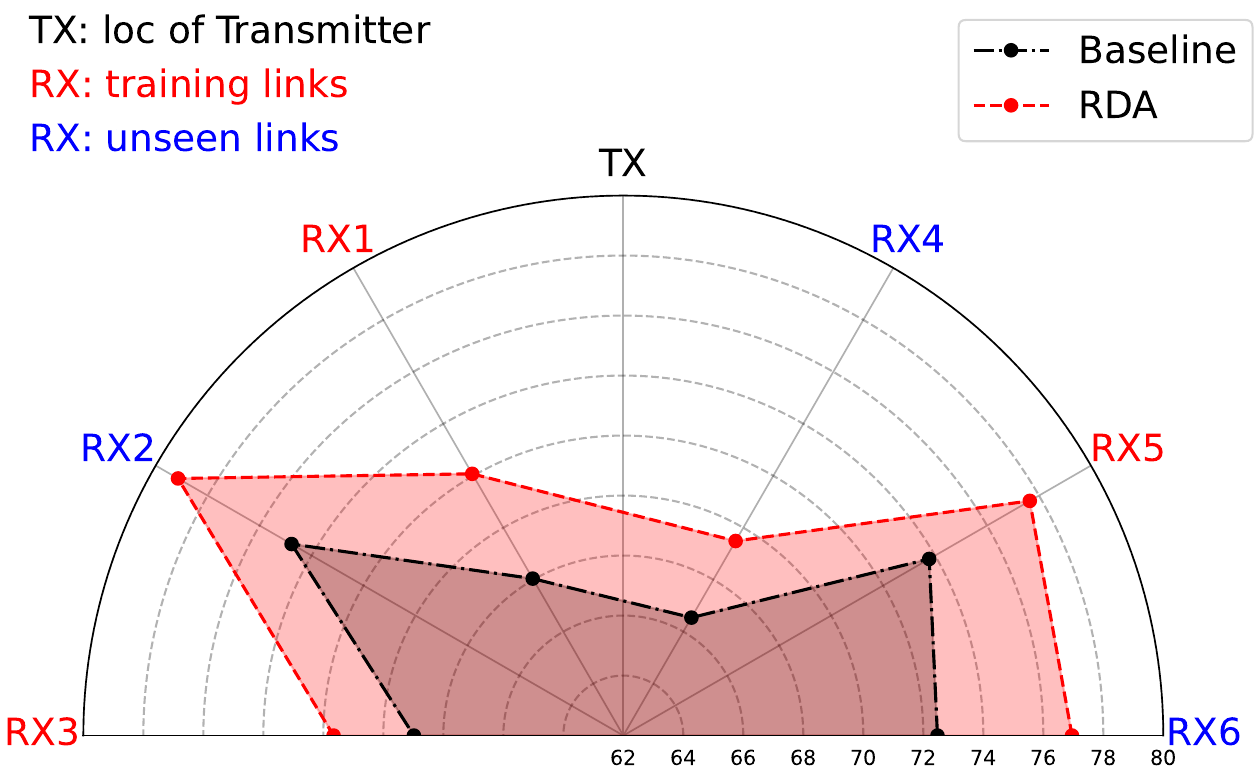}
    \caption{Performance across links in their actual locations.}
        \label{fig:xdev}
  \end{minipage}
  \hfill
  \begin{minipage}{0.465\textwidth}
    \centering
    \includegraphics[width=1\textwidth]{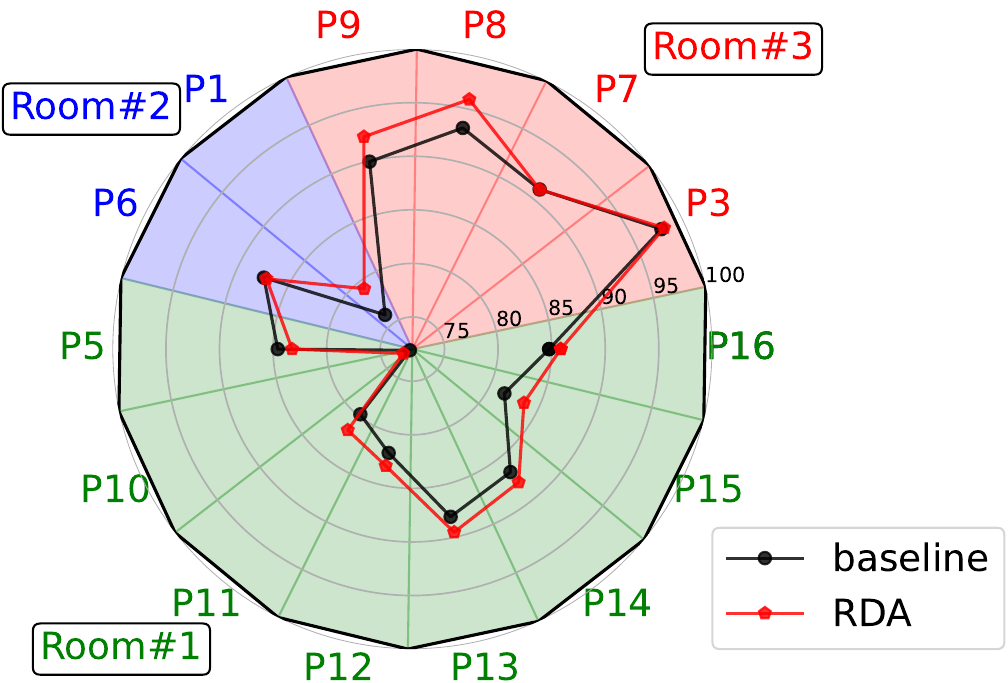}
    \vspace{-0.2in}
    \caption{Performance across Env.(Room) \& Persons (P$i$).}
    \label{fig:xperson}
  \end{minipage}
\end{figure}

\subsection{Cross-domain Generalization Ability}
\label{subsec:xdomain_ability}
We conduct both in-domain and cross-domain evaluation on Widar3 dataset using RF-Net as a baseline model.

\subsubsection{In-domain Performance}
Firstly, for in-domain evaluation, as in \fig\ref{fig:widar3-sota-x}, both training and test sets contain samples in the same domain. And RDA achieves an average performance gain of 4.23\% over all models. 

\subsubsection{Cross-Rx Evaluation} 
\rev{As mentioned above, \sysname improves cross-Rx performance by an average of 5.4\% on 9 models.
We now dive into the performance on each Rx device when training on Widar3-G6 and using RF-Net}. Specifically, cross-Rx evaluation constructs a training set with 3 out of 6 Rx devices and evaluates on 3 unseen Rx devices. As depicted in \fig\ref{fig:xdev}, the result shows improvement for all Rx devices, with an average performance gain of 3.6\% and peaking at 4.5\% on Rx6.

\subsubsection{Cross-Room and Cross-Person Evaluation}
\rev{In addition, we conduct cross-room and cross-person evaluation on Widar3-G6 dataset. The training dataset comprises four individuals in Room\#3, and the model's generalization ability is subsequently assessed by testing on other unseen rooms and individuals. As depicted in \fig\ref{fig:xperson}, the evaluation highlights an average accuracy enhancement of 0.87\% in cross-person scenarios and 1.1\% in cross-room scenarios. Notably, Person\#17 exhibited atypical performance, and thus, 
for the sake of clarity in the graphical representation, their data was excluded from the visualization
(still included in performance analysis), where the test result of baseline is markedly lower at 46.45\%, compared with the average performance of 85.36\% among all other individuals.}

\revc{Notably, we observed that the baseline performance in Cross-Room and Cross-Person evaluations is approximately 10\% higher than in the Cross-RX evaluation. However, the improvement observed in these cross-domain evaluations is not as significant as that seen in the Cross-RX evaluation. This discrepancy underscores the need for continued research into robust and adaptable techniques.
}

\subsection{Microbenchmarks}
\label{subsec:microbenchmarks}
\subsubsection{Ablation Study}
\label{sssec:ablation}
\begin{figure}[t]
  \begin{minipage}{0.48\textwidth}
    \centering
    \includegraphics[width=0.8\textwidth]{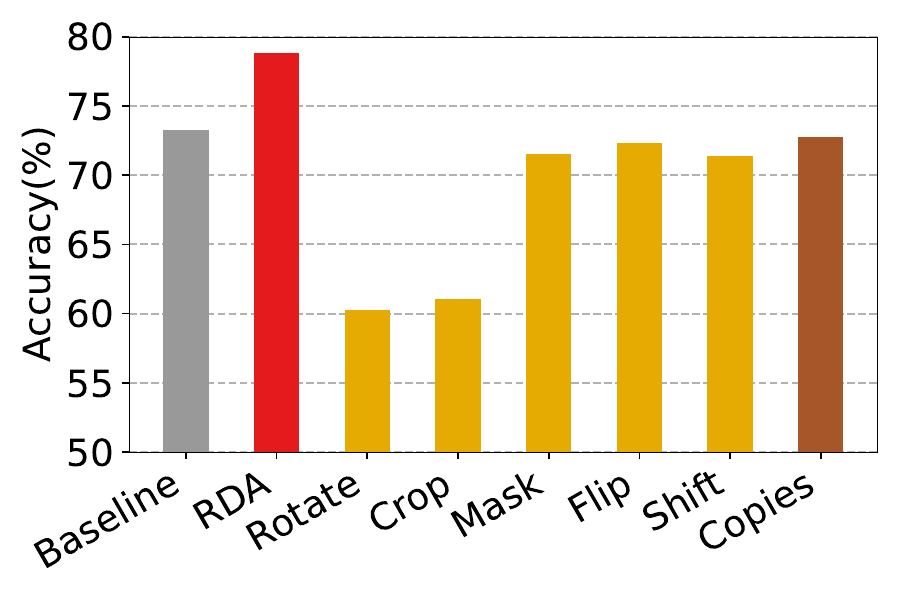}
    \vspace{-0.1in}
    \caption{Comparisons between IDA and RDA.}
    \vspace{0.5in} %
    \label{fig:ida_vda_results}
  \end{minipage}
  \hfill
  \begin{minipage}{0.48\textwidth}
    \centering
    \includegraphics[width=0.8\textwidth]{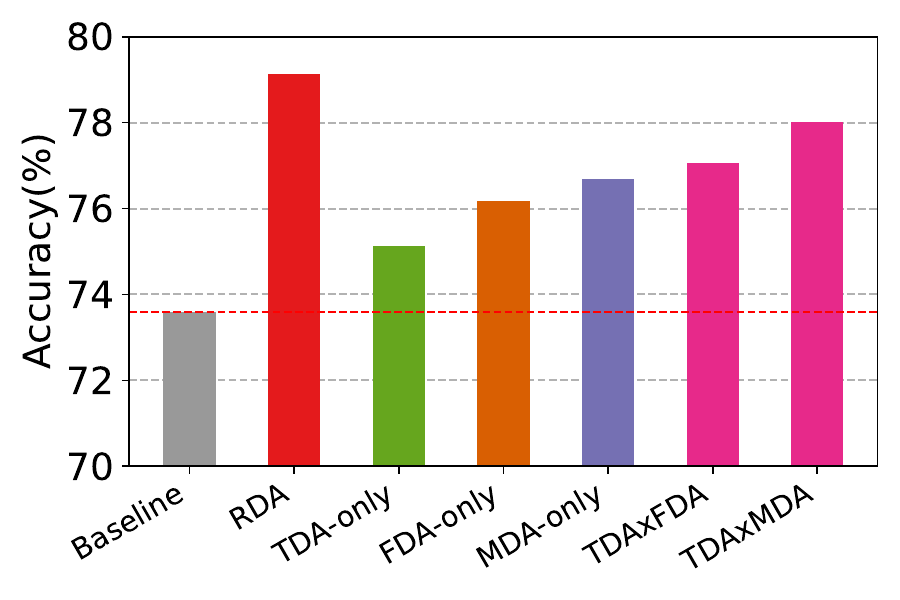}
    \vspace{-0.1in}
    \caption{\rev{Ablation study on different RDA policies.}}
    \vspace{0.5in} %
    \label{fig:ablation}
  \end{minipage}
  \hfill
\end{figure}

\rev{Unless stated otherwise, we use RF-Net as the baseline model and focus on cross-Rx evaluation with the Widar3 dataset for the study hereafter.
We conduct an ablation study to evaluate the effectiveness of TDA, FDA, and MDA, respectively. 
As depicted in \fig\ref{fig:ablation}, the overall performance gain reaches 5.52\%, achieved by RDA (ARatio=9, 2$\times$TDA, 6$\times$MDA, 1$\times$GSM). We further decompose RDA into three major modules of RDA, denoted as \textit{TDA-only}, \textit{FDA-only}, and \textit{MDA only}, respectively. Moreover, their combinations are also studied, including TDA$\times$FDA and TDA$\times$MDA. The result shows standalone policies, \textit{TDA only}, \textit{FDA only}, and \textit{MDA only}, can improve the performance by 1.54\%, 2.59\%, 3.1\%, respectively, while their combination of TDA$\times$FDA and TDA$\times$MDA, which leave out MDA or FDA, achieve performance gain of 3.48\% and 4.43\%. Among all three major modules, TDA contributes the least while its combination with other modules still adds to gains by around 1\%. 
We also observe that the performance improvement achieved through the introduction of GSM is not proportional to its high computational demand. By default, GSM requires the calculation of spectrograms for all subcarriers. Considering any budgetary constraints, GSM can be omitted with only a minimal impact on accuracy, specifically a decrease of 1.09\%, thereby avoiding the computation of unnecessary spectrograms, saving both storage and computational resources.
}

\begin{figure}[t]
  \begin{minipage}{0.48\textwidth}
    \centering
    \includegraphics[width=0.8\textwidth]{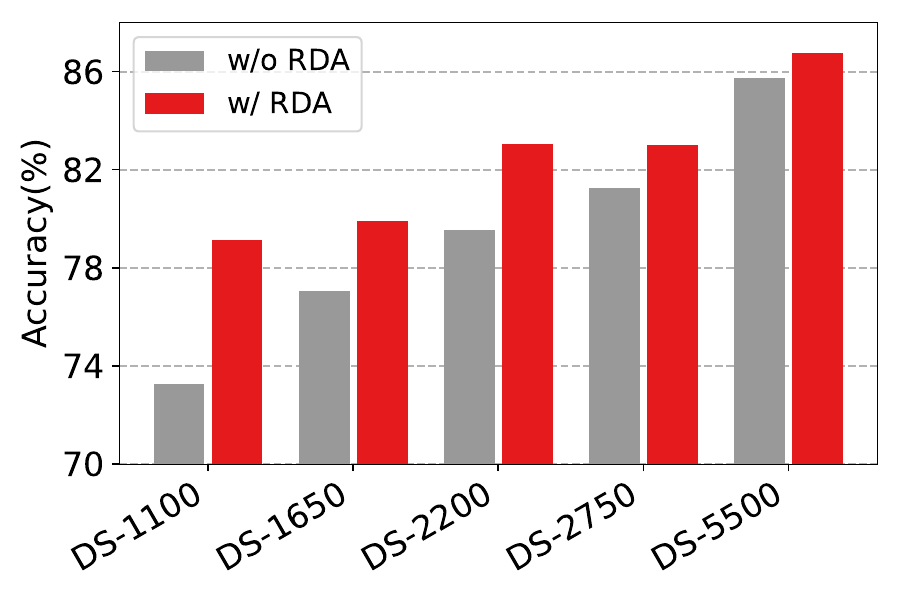}
    \vspace{-0.1in}
    \caption{Impact of data size.}
    \label{fig:exp-size}
  \end{minipage}
  \hfill
  \begin{minipage}{0.48\textwidth}
    \centering
    \includegraphics[width=0.8\textwidth]{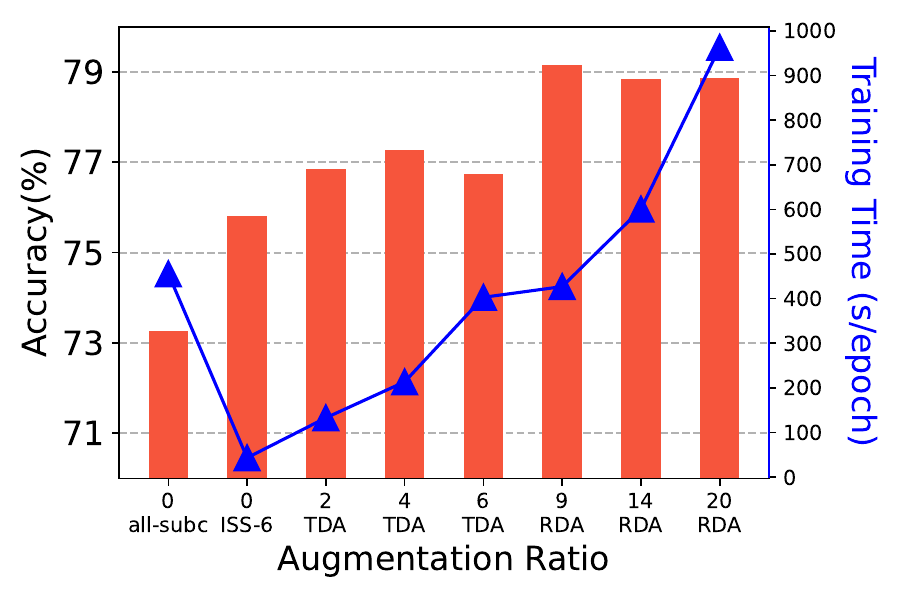}
    \vspace{-0.1in}
    \caption{Impact on Augmentation Ratio.}
    \label{fig:aratio}
  \end{minipage}  
\end{figure}

\subsubsection{Comparison with Image Data Augmentation (IDA)}
\label{sssec:compare_ida}
We undertake a comparative study to establish \sysname's superiority over existing IDA techniques in DWS, adhering to the same augmentation ratio of 5 for a fair comparison. RF-Net serves as our baseline.
Results in \fig\ref{fig:ida_vda_results} indicate that while certain IDA methods, like masking, slightly boost accuracy, others such as rotation are detrimental. Duplicating samples yields no benefit. In contrast, RDA outperforms IDA, not only enhancing performance but also offering physical interpretability. While this doesn't negate the applicability of all IDA techniques for DWS, it underscores the distinct advantage of RDA.

\rev{
\subsubsection{Comparison with SpecAugment}
\label{sssec:compare_specaug}
We refer to an open-source repository\footnote{https://github.com/zcaceres/spec\_augment} for the implementation of SpecAugment\cite{Park2019SpecAugmentAS}, an augmentation method for mel spectrograms of audio data, featuring time warping, time masking and frequency masking. Thus, there are three parameters, namely, time warp parameter ($W_{warp}$), frequency mask parameter ($F_{freq}$), and time mask parameter ($T_{time}$). All parameters set the boundaries for the random magnitude of data manipulations (warping or masking).
For time and frequency masking, we pad the masked region with the mean value of the whole spectrogram. 
Note that, these parameters have to be searched manually and we select them such that the size of masking and range of warping closely match in magnitude with those used (size of masking and range of shifting) in our proposed \textit{Motion-aware Data Augmentation}, where $W_{warp}=5$, $F_{freq}=20$, $T_{time}=20$. The result shows SpecAugment achieves a performance gain of 2.39\% in Cross-Rx evaluation on Widar3-G6 dataset, compared with 5.52\% using RDA and 3.1\% using \textit{Motion-aware Data Augmentation}-only. 
The difference between \textit{MDA} and the masking in SpecAugment is whether or not involving the information of motion in shifting distance decision and masking region selection. Additional gains when involving identified motion period also prove our design rationale in \S\ref{subsec:design_pda} that masking motion-specific region can alleviate overfitting towards local features and thus improve cross-Rx performance.
We also evaluate SpecAugment on Widar3-G22 dataset, as depicted in \fig\ref{fig:widar3-gesture-6vs22}. SpecAugment improves the performance by 2.67\% while RDA achieves 7.16\%. We further visualize the confusion matrices of both methods by analyzing how much gains they achieve compared with the baseline, shown in \fig\ref{fig:widar3_cm_gesture22}. We observe that although SpecAugment improves the performance, it seems to show a preference for the gesture types, such as Gesture4-9, and can even improve Gesture21 by 12.4\%. This implies SpecAugment only suits for specific gesture patterns. Differently, our proposed RDA works positively for most of these 22 gestures.
}

\subsubsection{Impact of Original Data Size}
Unlike most generative models, \sysname performs well with smaller training sets, but its performance varies depending on the data size. To examine the impact of training data size and assess the potential for reducing data collection efforts, we downsample the Widar3 training set to sizes of 1100, 1650, 2200, 2750, and 5500 and apply our RDA method while keeping the testing set consistent. \rev{The outcomes, as depicted in \fig\ref{fig:exp-size}, reveal that RDA empowers smaller datasets to reach performance levels equivalent to those achieved with datasets up to twice their size. This finding illustrates the efficacy of RDA in enhancing model performance through efficient data utilization, thereby providing a substantial avenue for addressing data scarcity and optimizing performance even when data availability is limited.}

\subsubsection{Augmentation Ratio}
In this experiment, we assess the influence of the augmentation ratio (ARatio), which is the ratio of the augmented to the original dataset size. \fig\ref{fig:aratio} reveals that higher ARatios improve performance but also incur greater computational cost and extended training time. For the Widar3 dataset, an ARatio of 9 provides a balanced trade-off between accuracy and computational overhead.

\subsubsection{Subc-as-channel v.s. Subc-as-sample}
CV-based DNN models often support multi-channel input, exemplified by PyTorch ResNet implementations \cite{pytorch_resnet} that accept input shapes of \( (C, H, W) \), where \( C \) signifies channel dimensions like RGB. For DWS models that consider spectrograms as input, certain DWS models draw inspiration from CV architectures, aligning subcarriers in the channel dimension. Conversely, the additional dimension in transitioning from CSI to spectrograms necessitates treating subcarriers as samples or aggregating them in models like RF-Net.

We assess these contrasting approaches using Widar3 trained on ResNet18. By consolidating all subcarriers into a single sample plus an augmented sample via \sysname (ARatio = 1), we find that Subcarrier-as-channel outperforms Subcarrier-as-sample by 6\%, with performance boosts of 1.83\% and 3.64\% when augmented with RDA, respectively. These findings advocate for treating subcarriers as channels in CV-inspired DNN models. Regardless of the input configuration, RDA enhances performance across the board.

\begin{figure}[t]
  \begin{minipage}{\textwidth}
  \centering
     \subfloat[]{%
          \includegraphics[width=0.38\textwidth]{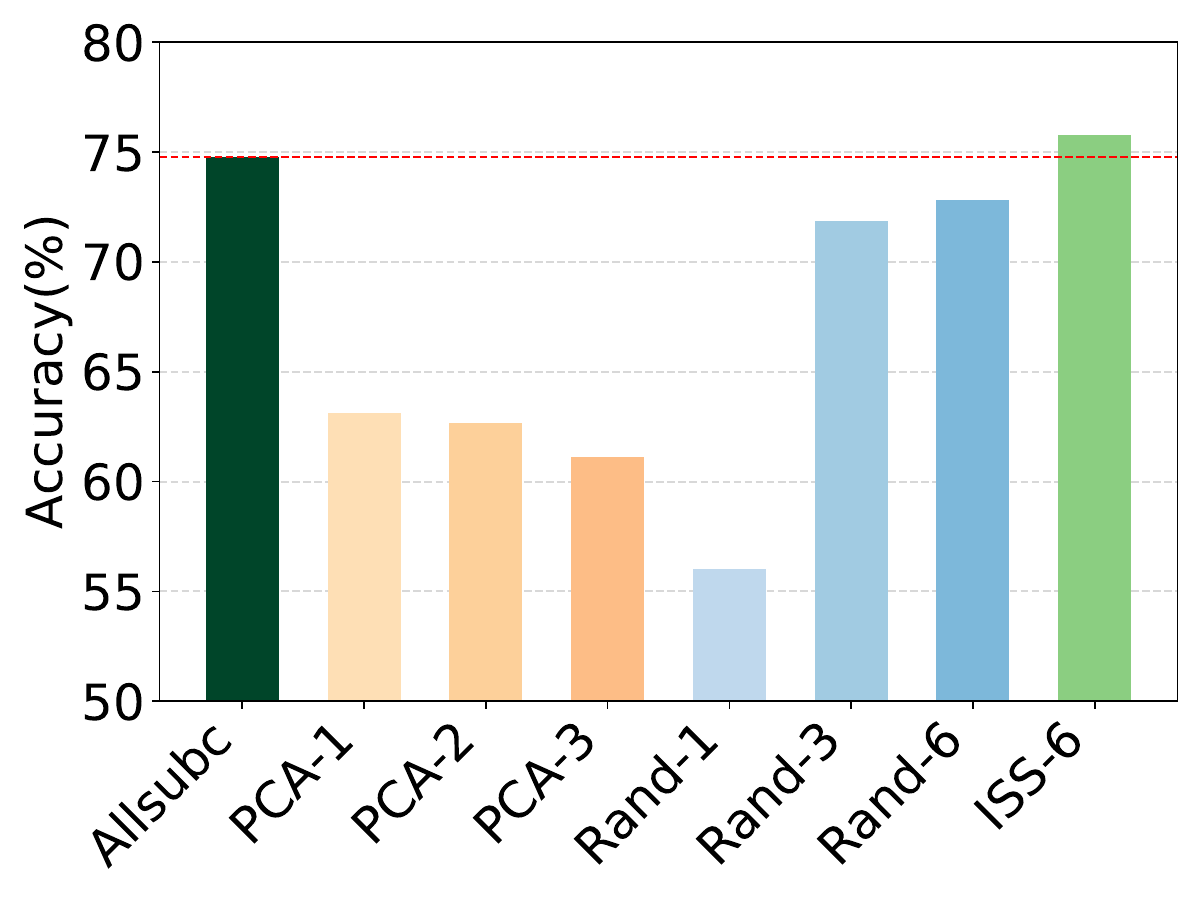}
            \label{subfig:pca_random}
    	  }
     \subfloat[]{%
          \includegraphics[width=0.25\textwidth]{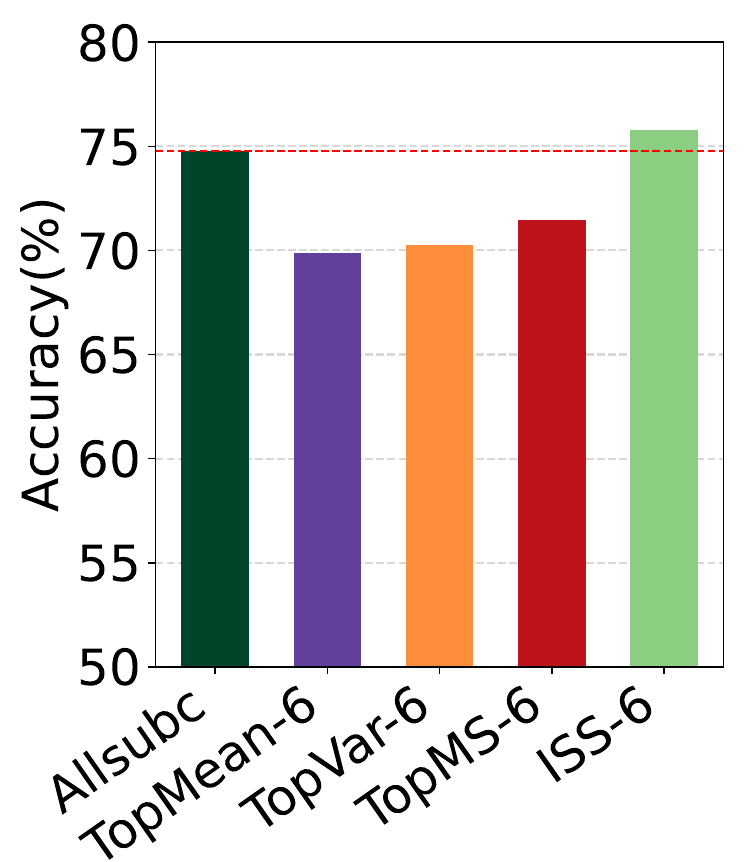}
        \label{subfig:mean_var_mstop_iss}
    	  }
     \vspace{-0.1in}
     \caption{Comparative analysis between using all subcarriers (Allsubc) and various metrics for subcarrier selection. (a) Allsubc v.s. top-$i$ principle components of all subcarriers (PCA-$i$), and randomly selected $i$ subcarriers (rand-$i$), (b) Allsubc v.s. subcarriers with top-6 mean value (TopMean-6), subcarriers with top-6 variance (TopVar-6), subcarriers with top-6 motion statistics (TopMS-6), and ISS-6.}
  \end{minipage}
\end{figure}

\subsubsection{Comparative Analysis of Subcarrier Selection Methods}
\label{sssec:compare_sel}
Within \sysname, Individual Subcarrier Selection (ISS) has emerged as a prudent approach for choosing representative subcarriers, significantly reducing training efforts. Although \sysname's primary focus is on data augmentation rather than selection, it prompts an intriguing question: 'How well do existing subcarrier selection methods perform in deep wireless sensing?' We aim to address this question from an experimental perspective:
Despite PCA's widespread adoption for effective dimension reduction while maximizing CSI variances in wireless sensing, its efficacy in deep wireless sensing is limited. In our study, we juxtapose Individual Subcarrier Selection (ISS) with PCA on the Widar3 datasets, employing the top-1, top-2, and top-3 principal components (PCA-1, PCA-2, PCA-3). Notably, PCA, when applied subcarrier-wise as in \cite{Zhang2021Widar30ZC, palipana2018falldefi}, results in accuracies of 63.1\%, 62.65\%, and 61.14\% respectively, which are approximately 9-10\% lower than using three randomly selected subcarriers (Rand-3), and significantly lower than ISS-3. As depicted in \fig\ref{subfig:ds-ratio} and \fig\ref{subfig:pca_random}, we observed that PCA's performance is similar to a single subcarrier selected based on motion statistics (ISS-1), a criterion that provides physical context which is absent in PCA.
In addition to PCA, we assess some traditional subcarrier selection methods like TopVar and TopMean in deep wireless sensing scenarios, which focus on maximum variance and mean amplitude, respectively. To better understand ISS, we examined a variant of ISS, termed TopMS, which selects subcarriers based on the highest motion statistics across the entire frequency band instead of among sub-bands as ISS does (See \S\ref{subsec:design_pda}). As illustrated in \fig\ref{subfig:mean_var_mstop_iss}, TopMS-6 outperforms all other intuitive methods but falls 4.32\% short of ISS-6. This is because ISS-6 leverages both physical characteristics and the diversity inherent in distinct sub-bands, resulting in a more comprehensive and motion-aware subcarrier selection method.

\subsubsection{Test-Time Augmentation Performance}
To evaluate the effect of test-time augmentation, we re-run the test phase, using the previously trained RF-Net on Widar3, by augmenting each test sample in the same way as the training data. For the same test sample, majority voting is applied to aggregate multiple copies generated by augmentation methods.
Our results show that test-time augmentation can bring an additional improvement of 0.7\%, further increasing the best RDA accuracy from 78.84\% to 79.54\%.
The computation overhead also increases marginally, adding an extra delay of 27 ms to each test sample with an ARatio of 9. 

\subsubsection{Computational Complexity}
We mainly focus on both the time and space complexity of \sysname. For time complexity, as mentioned in \S\ref{subsec:effectiveness}, we downsample all subcarriers before applying RDA, which greatly reduces the time for training the augmented dataset while achieving similar high performance as depicted in \fig\ref{fig:aratio}. We highlight that \sysname cache-and-fetch mechanism also saves approx. 50\% training time from the second epoch compared with 1st epoch for every augmentation method. 
We further analyze the space complexity, which apparently also increases as RDA creates more data.
Note that, all the computation overhead only occurs in the training stage, which is usually done offline and therefore less time-sensitive. The raw CSI data for Widar3 dataset occupies 78 GB, and the extra space usage increases linearly with the augmentation ratio.

\section{Discussions and Future Works}
\label{sec:discussions}

\sysname pioneers the study of radio data augmentation for deep wireless sensing. 
The experimental results show encouraging performance as well as inspire many follow-up research topics. There is certainly room for improvement and we discuss some limitations and future works here. 

\subsection{RDA Beyond Spectrograms} 
The current \sysname is mainly designed within the scope of WiFi sensing using spectrogram learning \cite{slnet}. 
We have already shown the proposed approaches can also be used for highly processed data representations like BVPs \cite{Zhang2021Widar30ZC,jiang2020towards}, which are based on DFS spectrograms. 
The proposed methods can also be directly applied to other forms of spectrograms, \eg, those generated by ACF calculation \cite{wu2020gaitway,yang2022rethinking}. 
Furthermore, we believe the idea of PDA can be extended to more data representations of WiFi. 
We are particularly interested in investigating and incorporating raw CSI augmentation in \sysname. 

\subsection{RDA Beyond WiFi} 
\label{ssec:beyond_wifi}
Another exciting topic is to exploit RDA wireless sensing beyond WiFi, \eg, millimeter wave (mmWave), UWB radars, etc. 
For example, in addition to time diversity, there are generally more antennas available for mmWave sensing, which underpin more space diversity. For FMCW mmWave radar, it is also possible to explore segmented frequency bands to increase frequency diversity \cite{jiang2020mmvib}. Both can be explored for data augmentation for mmWave sensing. 
Despite increasing efforts on data synthesis for mmWave sensing recently \cite{ahuja2021vid2doppler,bhalla2021imu2doppler,vishwakarma2021simhumalator}, physical data augmentation is still lacking. 

\subsection{Auto Augment} Our experimental results show that there is seemingly an optimal size for post-augmented data. 
How to automatically determine the optimal final dataset size and select the best augmentation policies is thus another interesting question to answer. 
Auto augment has attracted numerous efforts with remarkable advances in image data augmentation \cite{zheng2022deep}, which can inspire auto augment for RDA. 

A promising direction is to leverage 
diversity and affinity metrics \cite{Lopes2021TradeoffsID} to automatically determine the policies and augmentation size that maximize the two values, without going through the heavy training process.

\subsection{Synthetic Data Generation} We currently mainly focus on data warping augmentation, which transforms existing data for more samples. 
We note that \sysname is true data augmentation, rather than data selection, as it generates new samples that do not exist in the original dataset. 
Synthetic data generation based on deep learning, \eg, GANs, can also create synthetic virtual samples for training. 
There are some recent efforts in this direction by using adversarial learning \cite{jiang2020wigan}, diffusion models \cite{chi2024rf}, and/or resorting to a secondary modality like video data or IMU data \cite{bhalla2021imu2doppler,cai2020teaching, ahuja2021vid2doppler}. 
The proposed approaches in \sysname are complementary to these approaches and can be used jointly. 
Our future work will explore the combination of physical data augmentation and synthetic data generation, for which we are particularly interested in physics-inspired data generation.

\section{Related Works}
\label{sec:related_works}

This work targets a new problem of radio data augmentation for deep wireless sensing. We briefly review the related literature on wireless sensing and data augmentation techniques. 

\subsection{Deep Wireless Sensing} 
Traditional approaches for wireless sensing mostly rely on signal processing and/or conventional machine learning, which have seen years of development and achieved great performance for some tasks \cite{qian2018widar2,wang2015understanding,zhang2019widetect,wu2020gaitway,zhang2019smars,hu2021defall,feng2021lte,xie2020combating}. 
Recently, with the great success of deep learning in many applications, deep learning has been used in wireless sensing to achieve encouraging performance \cite{li2020wi,Zhang2021Widar30ZC,xiao2021onefi,ding2020rf,zheng2021more,chen2021movi,ozturk2021gaitcube,wang2020push,yang2022deepear}. 
These solutions typically adopt existing neural networks and take either raw data as inputs or convert them into the frequency domain or certain feature space. 
For example, Widar3.0 \cite{Zhang2021Widar30ZC} leverages CNN and RNN networks and estimates the body velocity profiles as a location-independent representation. 
RFPose \cite{rfpose}, RFPose3D \cite{rfpose3d}, and RFAvatar \cite{rfavatar} also use CNN models to capture body skeleton and mesh. 
EI \cite{jiang2018towards}, CrossSense \cite{zhang2018crosssense}, RF-Net \cite{ding2020rf} and Wi-Learner \cite{wi-learner}use more sophisticated networks such as adversarial learning, transfer learning, meta-learning, mainly to handle the issue of environment dependency. 
WiSIA \cite{li2020wi} employs conditional GAN to achieve WiFi imaging, which is also studied in \cite{jiang2020towards,wang2019person} with deep learning. 
Besides high accuracy, a particular focus of DWS solutions is to reduce the need for large-scale training data. For example, many of them aim to achieve one-shot learning or cross-domain learning \cite{ding2020rf,xiao2021onefi,gu2021wione, Zhang2021Widar30ZC}. 
Some works take a further step to customize the networks in the hope of better fitting time series data. STFNets \cite{yao2019stfnets}, which extends DeepSense \cite{yao2017deepsense}, and UniTS \cite{li2021units} are both customized for temporal-spatial learning with STFT operators. 
STFNets also exploits time diversity with multi-resolution windows, but not for the purpose of data augmentation. 

Different from these methods, \sysname approaches the problem of limited data from the root, \ie, the training data, and presents a data-space solution of radio data augmentation, which is orthogonal to the above solutions. 

\subsection{Data Augmentation}
Some solutions explore beyond wireless and exploit cross-modality sensing and data synthesis to mitigate the data scarcity issue in DWS. 
IMU2Doppler \cite{bhalla2021imu2doppler} uses IMU datasets to train a network for mmWave signals with a small amount of labeled data. 
XModal-ID \cite{korany2019xmodal} considers the problem of WiFi-based person identification from a given video footage. A more interesting solution is proposed in \cite{cai2020teaching} to translate online videos to instant simulated RF data for training DWS systems without massive RF data collection. 
Vid2Doppler \cite{ahuja2021vid2doppler} also synthesizes Doppler mmWave data from videos. 
Different from these methods, our proposed RDA aims to inflate existing training data without collecting more or resorting to a secondary modality. 

Data augmentation is only primarily touched in the field. For example, MoRe-Fi \cite{zheng2021more} rotates the raw signals in the I/Q-plane for augmentation. 
OneFi \cite{xiao2021onefi} also augments data by rotating the velocity distribution, which is only applicable to specific tasks. 
Image data augmentation techniques like noise injection, scaling, cropping, and image mixture are applied to RF spectrograms in \cite{zhang2021AugDLSTM} for human activity recognition. 
Techniques like generative adversarial learning have been used to synthesize virtual samples for WiFi \cite{Xiao2019CsiGANRC,jiang2020wigan,Yang2022WiImgPT,yang2022rethinking, li_crossgr_2021} or radar \cite{vishwakarma2021simhumalator,ozturk2022toward,xue_towards_2023} signals. 
Differently, \sysname systematically studies the problem of RDA and presents physical data augmentation, rather than generating virtual samples. 
Nevertheless, RDA is complementary to and can be stacked on top of existing cross-modality training or data synthesis methods. 

Data augmentation is a well-studied area in CV and is extended to other fields such as speech processing \cite{Park2019SpecAugmentAS}. 
A comprehensive overview is provided in \cite{shorten2019survey} for literature review of image data augmentation. 
Although most image data augmentation techniques do not perform the best on wireless data, they inspire our study of a counterpart problem of radio data augmentation and motivate our design of \sysname for deep wireless sensing.

\section{Conclusions}
\label{sec:conclusion}
This paper presents \sysname, the first radio data augmentation framework for deep wireless sensing in the scope of Wi-Fi sensing. 
We propose a set of physical data augmentation techniques that inflate existing training data significantly and effectively. 
We achieve so by leveraging the inherent data diversity of wireless signals. 
We implement \sysname and integrate it with the state-of-the-art DWS models and evaluate it on three different WiFi sensing datasets. 
The results show that \sysname achieves remarkable accuracy improvements on baseline models and datasets. 
Furthermore, by simply performing RDA with a common CV network, we can achieve comparable or even better performance than customized networks for RF signals with highly-processed inputs. 
\sysname pioneers the study and inspires follow-up research on radio data augmentation. We also plan to extend \sysname to more wireless data representations and other wireless sensing modalities, which we believe will become a standard component for DWS to overcome the problem of limited data. 

\section*{Acknowledgment}
This work is supported in part by NSFC under grant No. 62222216 and Hong Kong RGC ECS under grant 27204522 and RIF under grant R5060-19.

\bibliographystyle{ACM-Reference-Format}
\bibliography{refs, zotero}

\end{document}